\documentclass{sig-alternate}
\usepackage[utf8]{inputenc}
\usepackage{url}
\usepackage{float}
\usepackage[caption = false]{subfig}
\usepackage{graphicx}

\usepackage{csquotes}
\MakeOuterQuote{"}
\begin{document}
\newcommand{\rpg}{\textit{RPG }}
\newcommand{\mtl}{\textit{Truemetal }} 
\newcommand{\swz}{\textit{Swzone }}
\newcommand{\psy}{\textit{Psychlinks }}
\newcommand{\mattia}{{}}
\newcommand{\enoch}{{}}

\graphicspath{{img/}}

\CopyrightYear{2016} 
\crdata{978-1-4503-3196-8/15/04}  

\title{Community structure and interaction dynamics through the lens of quotes}

%
%
%
%
%

\numberofauthors{2}\author{
\alignauthor
Mattia Samory\\
       \affaddr{Univ. Padova, Italy}\\
       \email{samoryma@dei.unipd.it}
\alignauthor
Enoch Peserico\\
       \affaddr{Univ. Padova, Italy}\\
       \email{enoch@dei.unipd.it}
}

\maketitle
\begin{abstract}

This is the first work investigating community structure and interaction dynamics through the lens of quotes in online discussion forums. We examine four forums of different size, language, and topic. Quote usage, which is surprisingly consistent over time and users, appears to have an important role in aiding intra-thread navigation, and uncovers a hidden “social” structure in communities otherwise lacking all trappings (from friends and followers to reputations) of today's social networks.\end{abstract}



\section{Introduction}
\label{sec:intro}
We examine four online forums of different size, language and topic through the lens of \emph{quotes} -- excerpts from previous posts that a new post can cite. This is an aspect of discussion that holds a wealth of information and yet has not been extensively investigated so far. We begin in Section~\ref{sec:quotes} with a brief look at how quoting works, and how it differs from other "discussion enrichment" mechanisms such as cites, likes or replies. In Section~\ref{sec:related} we review the related literature, both on these better-explored mechanisms, and on online forums in general. After some details in Section~\ref{sec:data} about our dataset and how we harvested it, in Section~\ref{sec:stats} we focus our attention on a number of basic quantitative metrics characterizing quotes in the four forums. Quote usage, albeit different in different forums, appears remarkably, almost eerily consistent across time and users in each forum. Also, although quotes share many of the "typical" characteristics of social environments such as heavy-tailed distributions, they markedly lack "rich-get-richer" characteristics. In Section~\ref{sec:nav} we explore the relationship of quotes with the distance (in time and post thread order) that separates quoted and quoting post; one interesting finding is that quotes appear, among other things, to play a crucial role in aiding thread navigation.
Finally, in Sections~\ref{sec:social} and~\ref{sec:applications}, we examine the implicit network that quotes effectively create between users -- in a context (that of online forums) that tends to lack, or see little use of, all the explicit trappings of modern social network, from likes to followings to friendships. We show that the quote network is effectively an "implicit" social network, and that each poster sports his own, characteristic interaction pattern, through which he can often be identified even when no other information except his local quote network is known.

\section{Quotes}
\label{sec:quotes}
\begin{figure}[h]
\includegraphics[width = \linewidth]{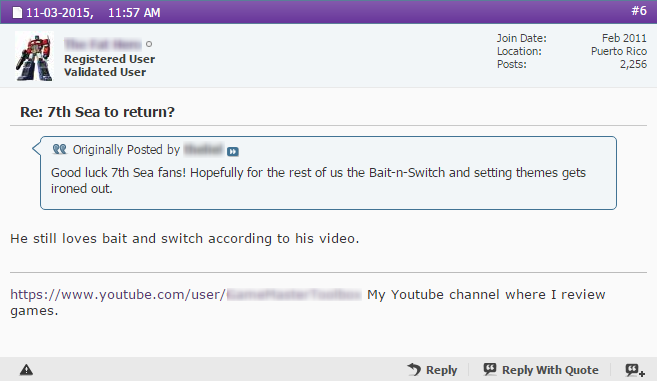}
\caption{Example post containing a quote from RPG. At the bottom right corner one can see various options for adding a new post: reply to this post, quote this post, and quote this post along with multiple other posts.}
\label{fig:post}
\end{figure}

Most online forums today offer a quotation mechanism, that allows a post author to cite excerpts of other posts -- both in the same and in other discussion threads. To do so, one simply clicks on a ``quote'' button that appears on the post to be quoted. This brings the entire quoted post, highlighted and preceded by ``Originally posted by <quoted author>'', into the new post at the current text insertion point. The new post's author then can manually edit the quoted post, and typically does so to remove less relevant passages (see Figure~\ref{fig:post}).

\vspace{0.5cm}

We remark that quotes are a widespread mechanism in forums, that differs from \emph{replies} (analysed e.g. in \cite{Aumayr2011}). A forum with replies links each post beyond the first to \emph{exactly} one previous post \emph{in the same thread} as a reply, effectively organizing the thread into a tree of posts rather than into a linear sequence. Unlike replies, quotes allow a post to link multiple previous posts (or none), potentially belonging to other threads or even subforums. Furthermore, quotes explicitly identify the portion of the linked post to which they refer -- in this sense they are more informative than simple citations like those of scientific citation graphs.

We can then see quotes as "higher resolution" versions of affordances such as retweets, shares, and cites, in the same roles as discussion aggregators and signals of attribution, acknowledgement, and endorsement
. Furthermore, although online forums are a less ``fashionable'' research venue compared to more modern platforms such as social networks, we believe that they can provide a picture of user interactions not only of higher resolution, but also with less noise, since they do not sport the same extensive level of automated, personalized, continuous curation of content -- a process that makes it often difficult to recover many details of the actual interaction (for example, exactly how a contribution that a user ``liked'' was shown to that user). Thus, we believe that the analysis of quotes in online forums can yield profound insights on the dynamics of online discussions -- insights that can also apply to, but would be harder to obtain from, more modern platforms.

\section{Related Work}
\label{sec:related}
This paper extends some of our preliminary research~\cite{Samory2015} that examines the role of quotes in coagulating and organizing discussion, and also suggests they could reveal the social structure of the debating community. We can divide related literature into three areas: discussion organization, evolution and interpretation; user identification and characterization; and emergence of social structure from interaction.

\subsection{Discussion analysis}
Considerable effort has been devoted to understanding how online discussion initiates, evolves, and is received by users. Conversation thread structure has been investigated mostly through patterns of post \textit{replies}, rather than \textit{quotes} \cite{Kumar2010,Aumayr2011}; interestingly, information on timing and user identity allegedly improves accuracy in reconstructing thread structure, which suggests online discussion is governed by social conventions richer than simple turn taking. An increasingly popular topic is that of predicting the propagation of a piece of content through retweets \cite{Kwak}, rumors \cite{Friggeri2014}, and memes \cite{Leskovec2009}; although these citation mechanisms resemble quotes in affording information sharing and source attribution, they are embedded within the frame of social and news media, platforms not designed for peer discussion. Looking at citation content instead of dissemination, recent research has built tools to interpret public dialog through quotes, exposing e.g. the systematic bias in news media outlets \cite{Niculae2015}, or what influences credibility in social media text \cite{Soni2014}.

\subsection{User identification and characterization} Quoting involves choices in when, whom, what, and how to quote -- choices that are part of a personal writing style and can thus help reveal information about the author. Previous authorship attribution efforts have used the \textit{presence} of quotes, alongside other linguistic and structural features, to identify the authors of a message in online forums \cite{Zheng2006} and email \cite{Vel2000}; our work, and in particular the results of Section~\ref{sec:applications}, differs in that it uses citation links in a \textit{corpus} of messages rather than in an individual message, and uses it as the \textit{exclusive} source of information. An approach closer to ours is that of Govidan et al. \cite{Govindan2013}, who have attempted deanonymization through network analysis -- but with the goal of restricting the number of lookalike nodes in the network, rather than of giving sharp identification within a pool of candidate users. 

A problem related to, but different from, identification is that of user \emph{characterization}: for example, retweets have been used to infer the “Big Five” personality traits of the users \cite{Adali2012}. While we believe the analysis of quotes could be profitably applied to this task, it is a line of research beyond the scope of this paper.

\subsection{Implicit networks} An extensive body of research has focused on understanding the social mechanisms triggering the creation of an edge in a social network -- both at the link level~\cite{Liben-Nowell2007, Hutto2013}, and at the entire network level~\cite{Leskovec2008}. However, it is an open question whether links in online social networks are reliable indicators of bonding.

A related research line of considerable practical interest involves inferring social networks from the actual observed interactions (such as exchanged messages or co-presence at events) \cite{Zhou2014}. Platforms analyzed in the literature include academic citation networks \cite{Leskovec2005}, online college communities \cite{Panzarasa2009}, email \cite{Roth2010}, and phone call logs \cite{Gupte2012}; yet quotes in online forums have never been investigated to date.


\section{Dataset}\label{sec:data}
\begin{figure}
\subfloat[]{\includegraphics[width = .5\linewidth]{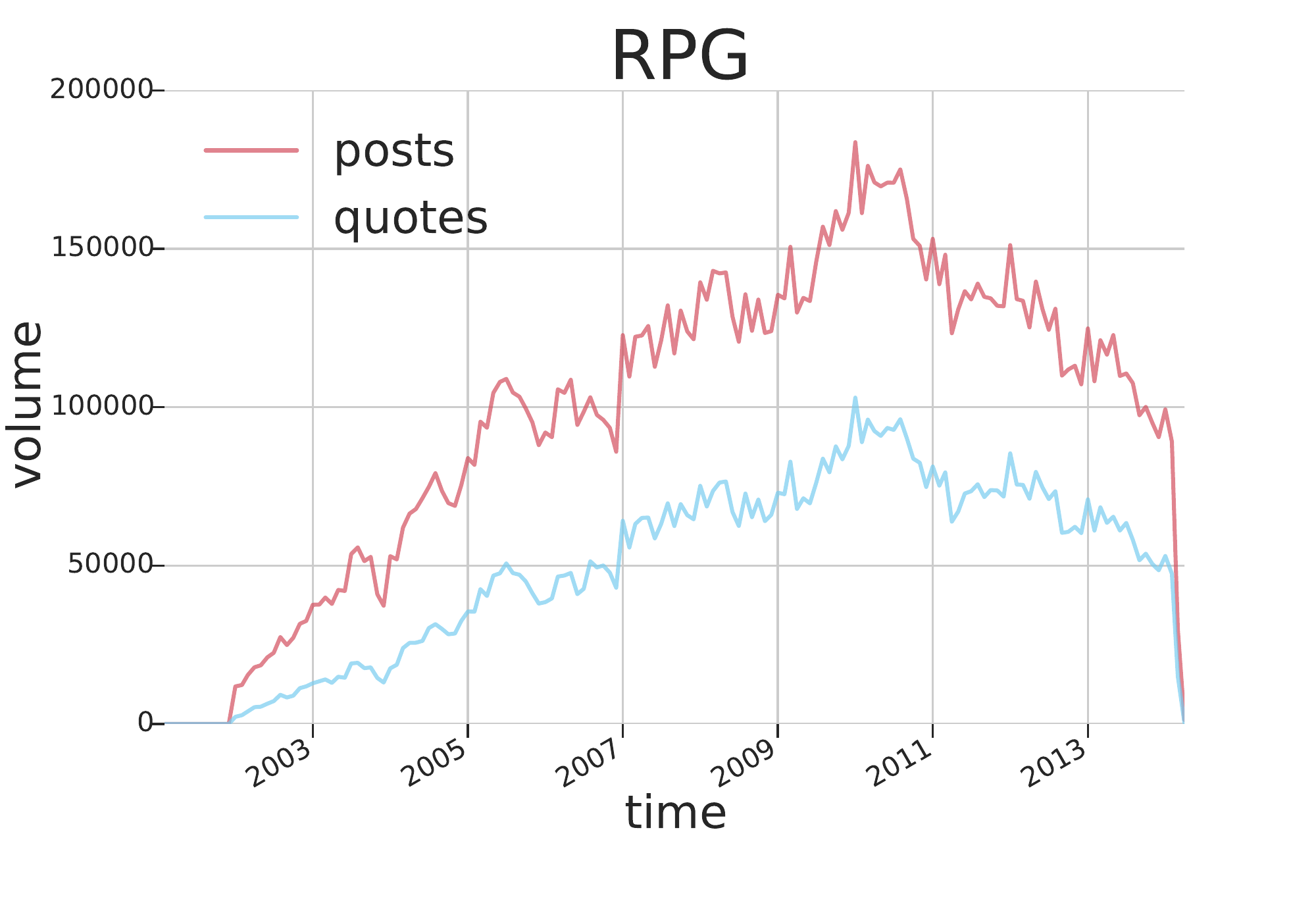}} 
\subfloat[]{\includegraphics[width = .5\linewidth]{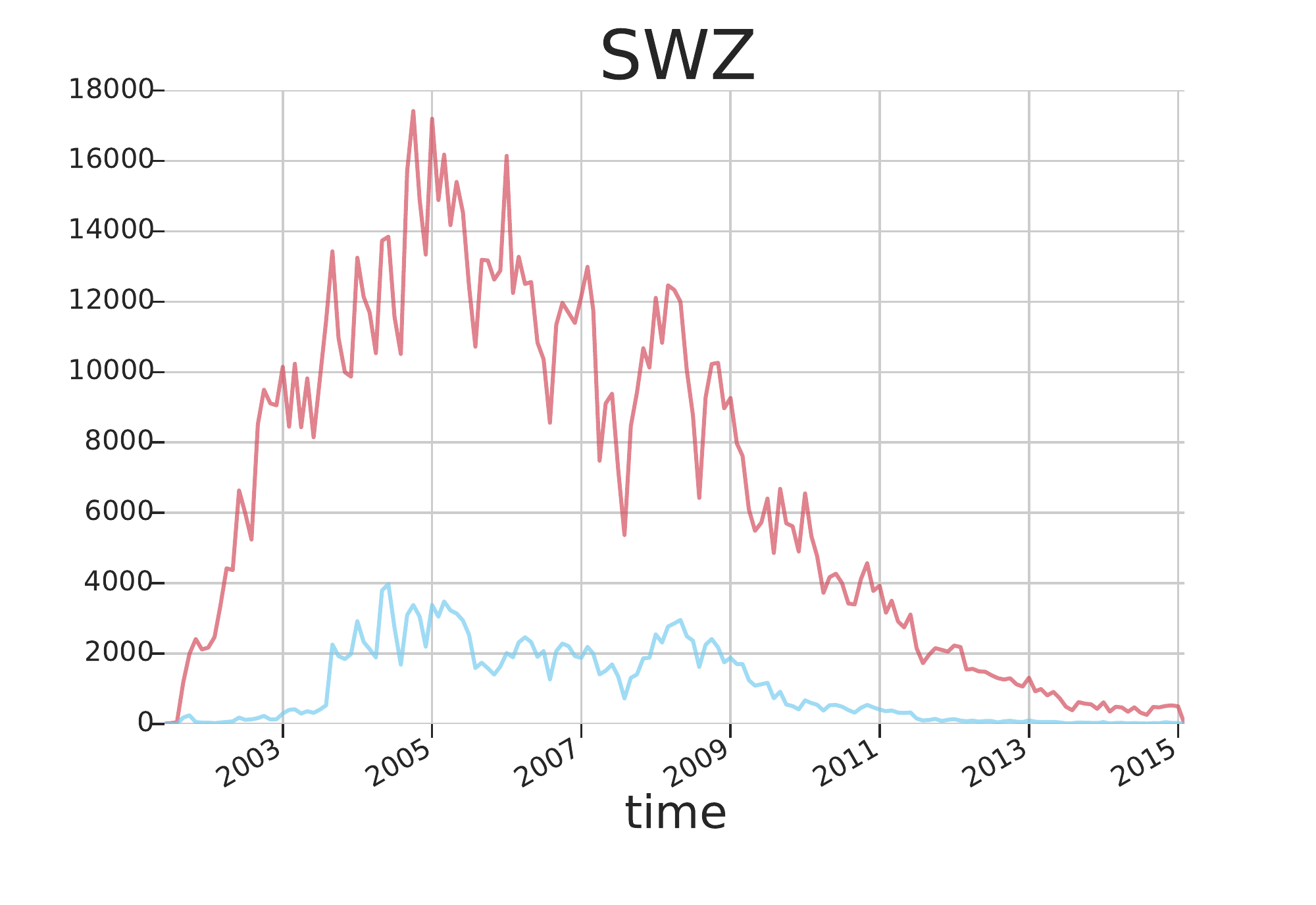}}\    \subfloat[]
{\includegraphics[width = .5\linewidth]{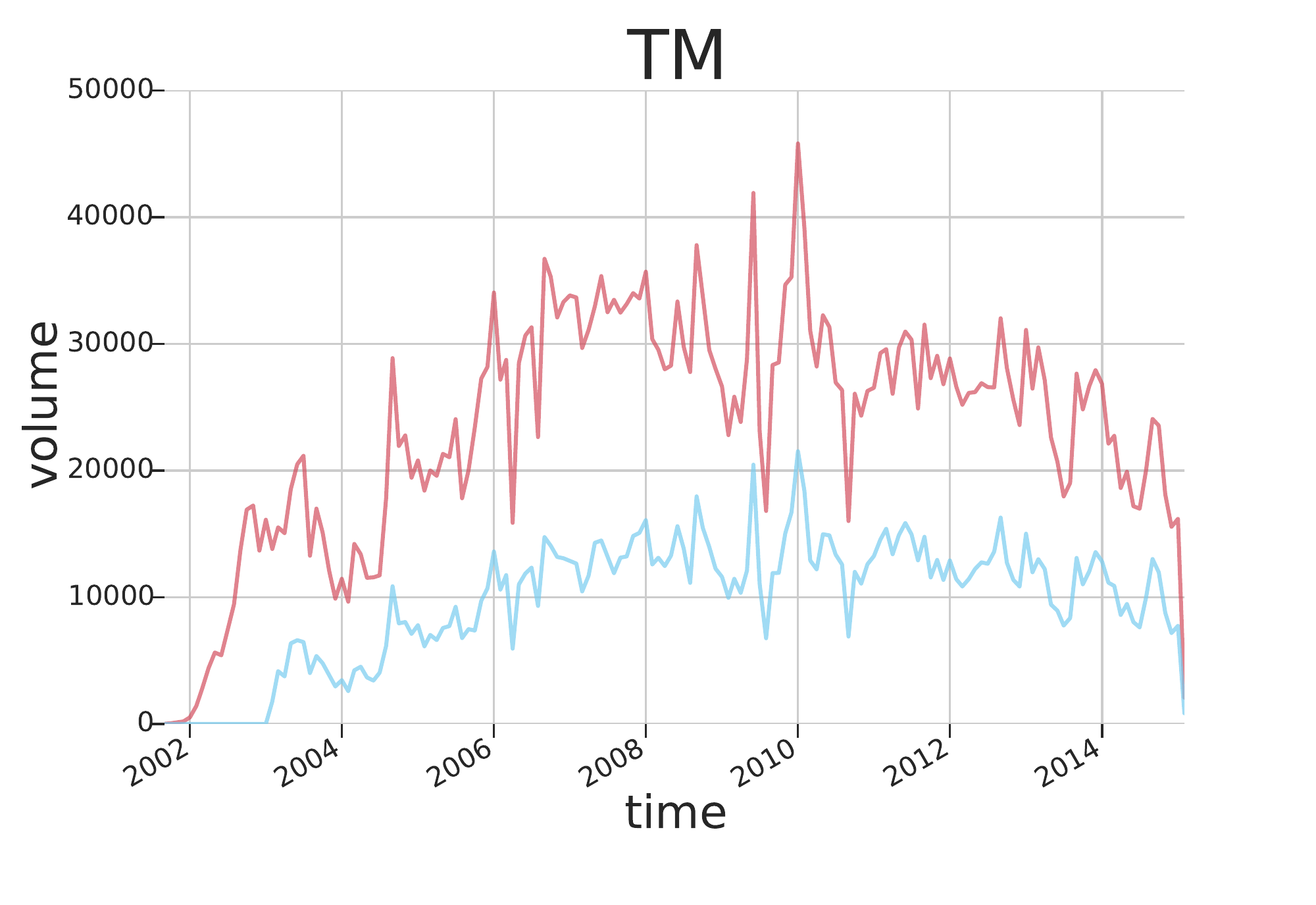}}
\subfloat[]{\includegraphics[width = .5\linewidth]{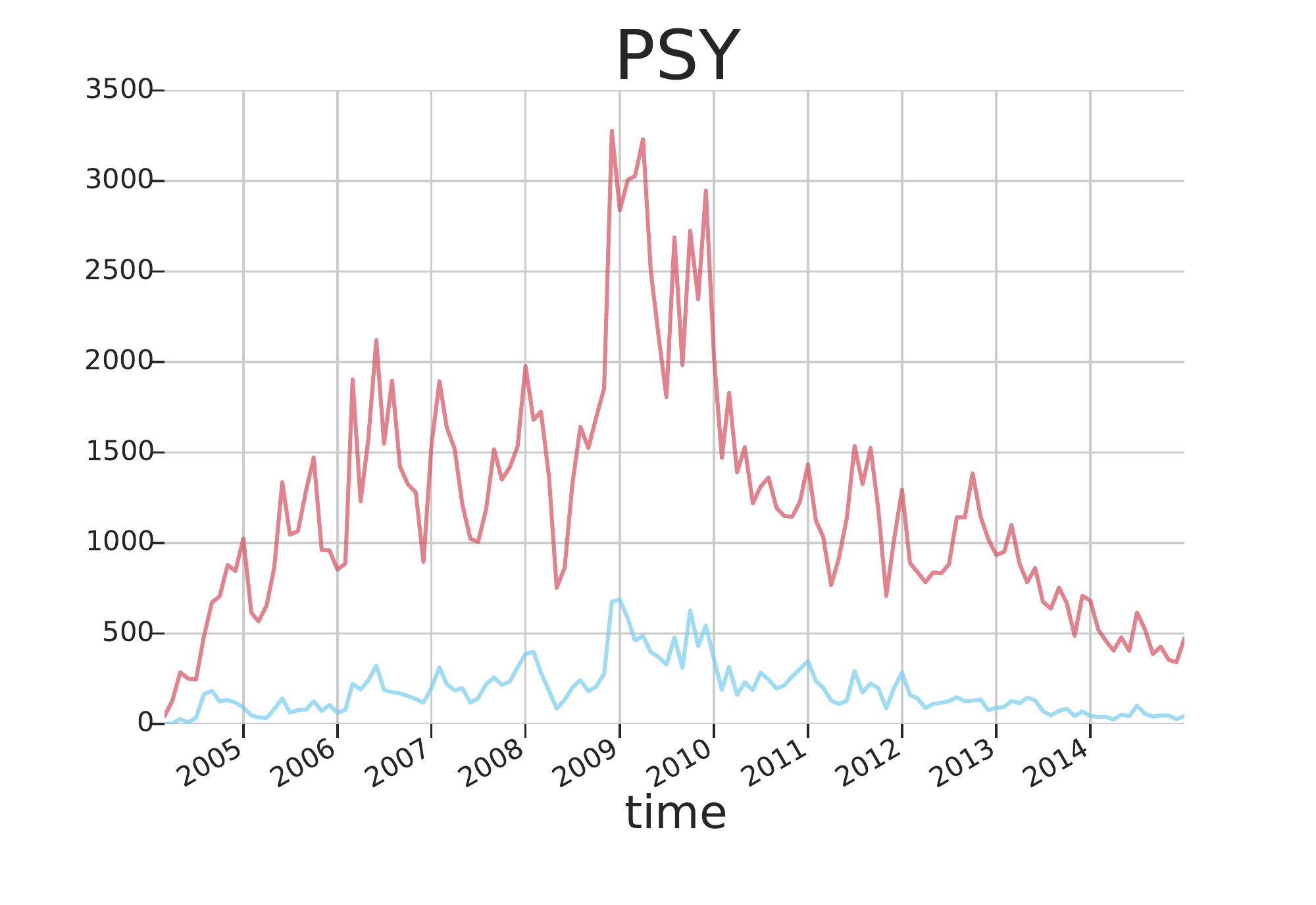}} 
\caption{Quote and post volume per month}
\label{fig:volume}
\end{figure}
We examine quotes in a range of diverse online forums, so as to provide a sense of how generalizable our findings are. This section briefly presents the four forums involved in the study, summarizes how data was gathered, clarifies the limits of the dataset, and discusses the difficulties encountered in the process.

\subsection{Four forums}
We examine four different forums so as to minimize bias from factors of scale, user background, and other patterns specific to an individual community.

\begin{description}
\item[RPG] is the largest international online forum devoted to roleplaying games (\textit{RPGs}), with a focus on tabletop rpgs. Its users come from many different backgrounds, and include a sizeable minority of professional game developers. The forum is divided into subforums that span a wide range of rpg-related topics, from speculations on new releases to \textit{play-by-post} online games.\footnote{\url{http://forum.rpg.net}}
\item[Swzone] is the forum section of an Italian IT news and information website. It is serves as a place for knowledge exchange between IT experts and the general public, and its threads feature user-contributed guides, problem troubleshooting, as well as software/hardware reviews.\footnote{\url{ http://forum.swzone.it}}
\item[Truemetal] is a major Italian board for discussing metal and hard rock music. Beside areas for casual conversation and music-related classified ads, most conversation revolves around critique of artists and albums, organized in subforums that reflect a taxonomy of subgenres. The community is active and engaged, and encourages users to meet in real life at concerts.\footnote{\url{ http://truemetal.it/forum}}
\item[Psychlinks] defines itself as a mental health support community. It gives information on the matters of psychology and personal development. Conversation usually happens in the form of comments to either an article on a particular condition, or personal stories. The forum, in English, is heavily moderated.\footnote{\url{http://forum.psychlinks.ca}}
\end{description}

It is clear how each forum specializes in a distinct topic. Table \ref{tab:size} shows how the forums differ by post, thread, and user cardinality. Two of the forums employ English as their main language (\textit{RPG, Psychlinks}), the other two Italian (\textit{Truemetal, Swzone}). The goal, focus, and typical evolution of discussion varies considerably across different forums -- and indeed even within each forum. For instance, \textit{RPG} features some subforums (topical subdivisions of a forum) dedicated to Q/A, others to review and commenting, and others still to conversation between peers; these are only some of the discussion patterns emerging from the datasets.

\subsection{Data gathering}
We crawled the four forums, acquiring all posts available since each forum's inception to the day of the crawl. We developed a python script to emulate what a freshly registered user would see logging into the forum, processing the current page top-to-bottom, and browsing to the next. The crawler proceeded breadth-first through the forum structure, first analysing subforums and saving links to threads, and then fetching posts from each thread. This procedure does not yield a perfect snapshot of the forums, as some posts contributed after the start of the crawl might have been included; however, such inaccuracies are extremely minor, since the crawl of even the largest forum required only a few days and for all four forums the number of posts per day is extremely small compared to the total post count (see Table \ref{tab:size}).

At the time of the crawl all four forums were using different customizations of the popular vbulletin\footnote{\url{https://www.vbulletin.com/}} community software. We parsed the scraped html pages through similar scripts, extracting metadata about users, subforums, threads, posts, and quotes, as well as the html tags surrounding the individual posts and quotes. All information acquired during the crawl was immediately stored into a database, and data curation was finalized at a later time.

\subsection{Quote curation}
\begin{figure}[t]
\subfloat[]{\includegraphics[width = .5\linewidth]{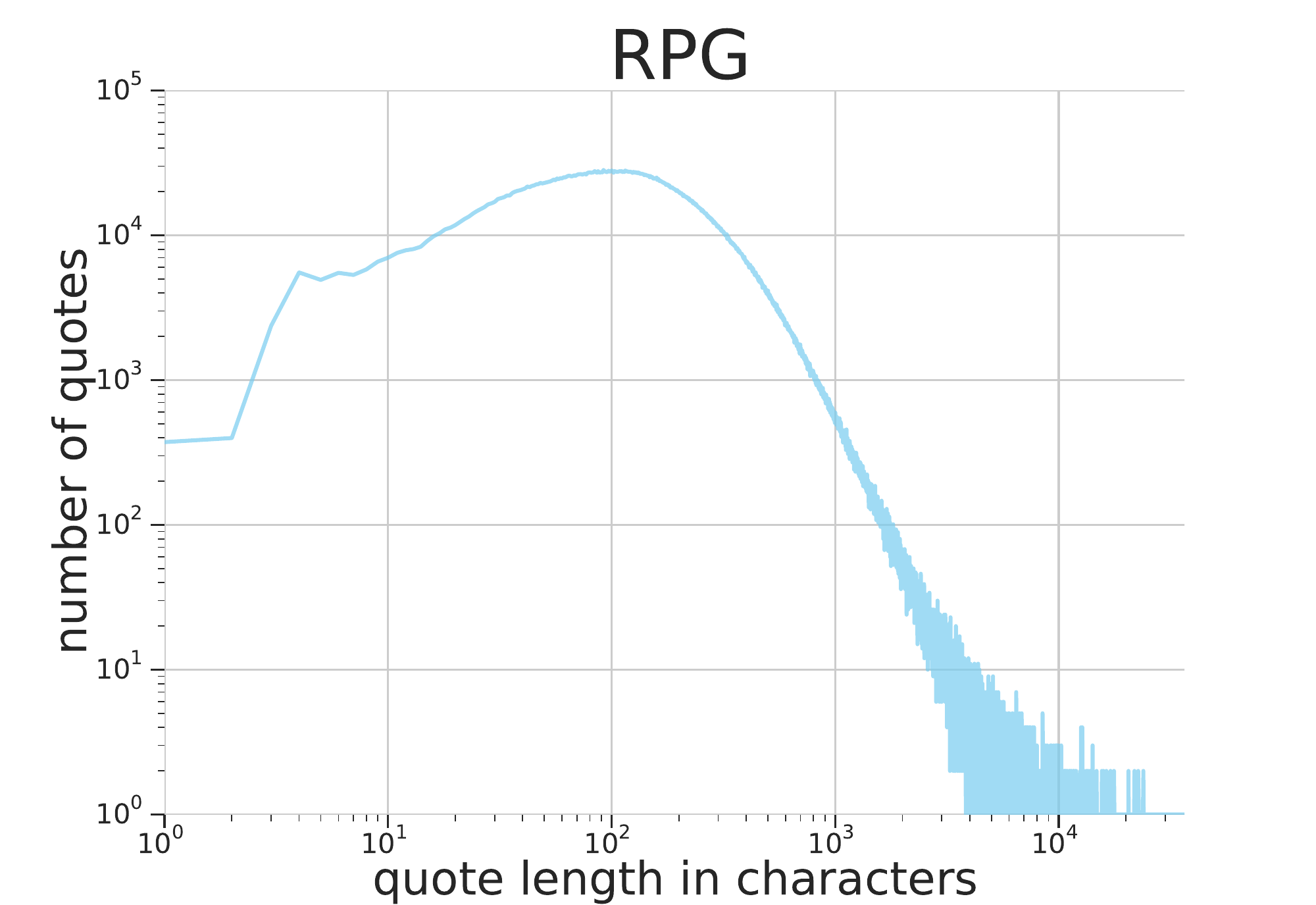}} 
\subfloat[]{\includegraphics[width = .5\linewidth]{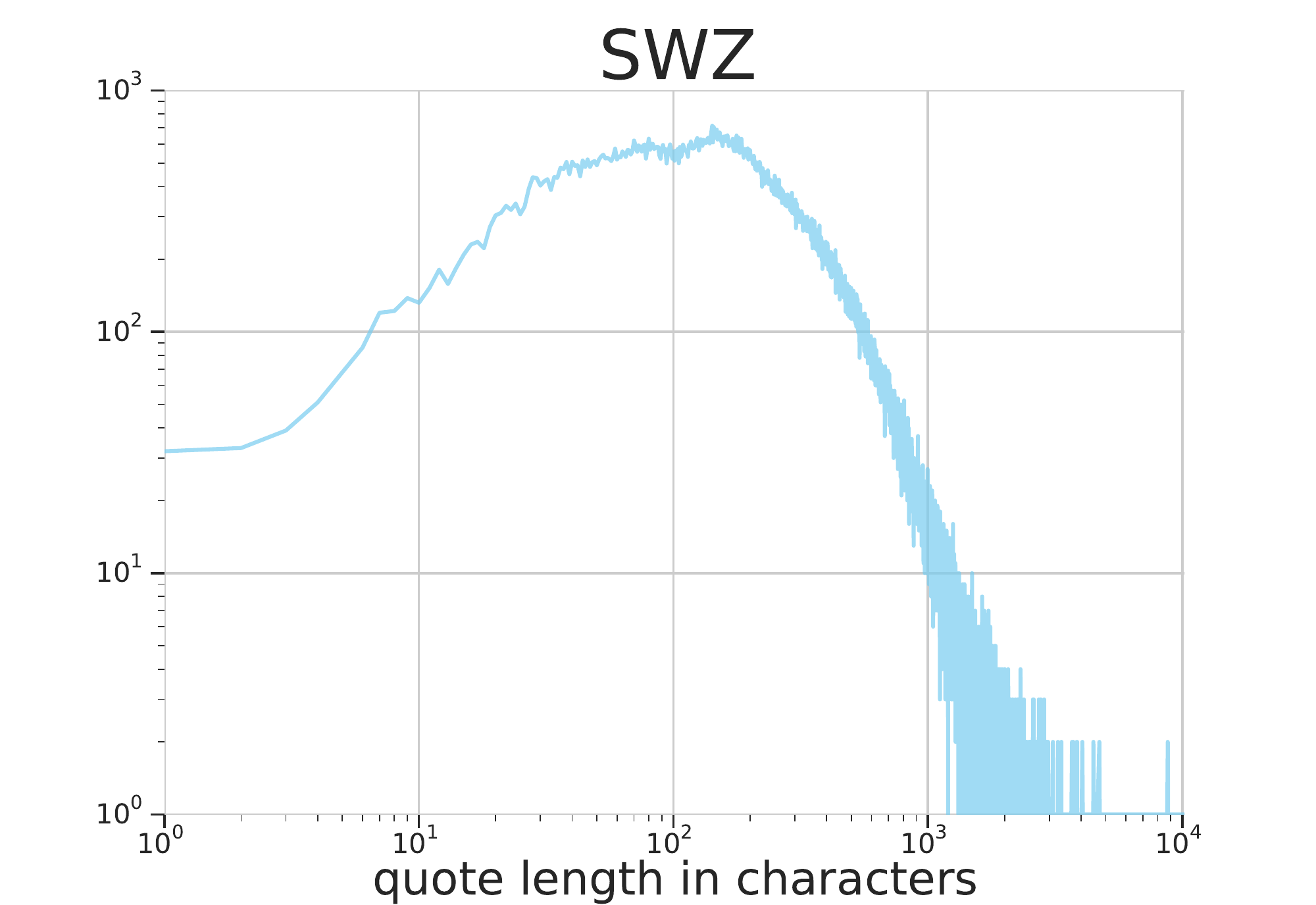}}\    \subfloat[]{\includegraphics[width = .5\linewidth]{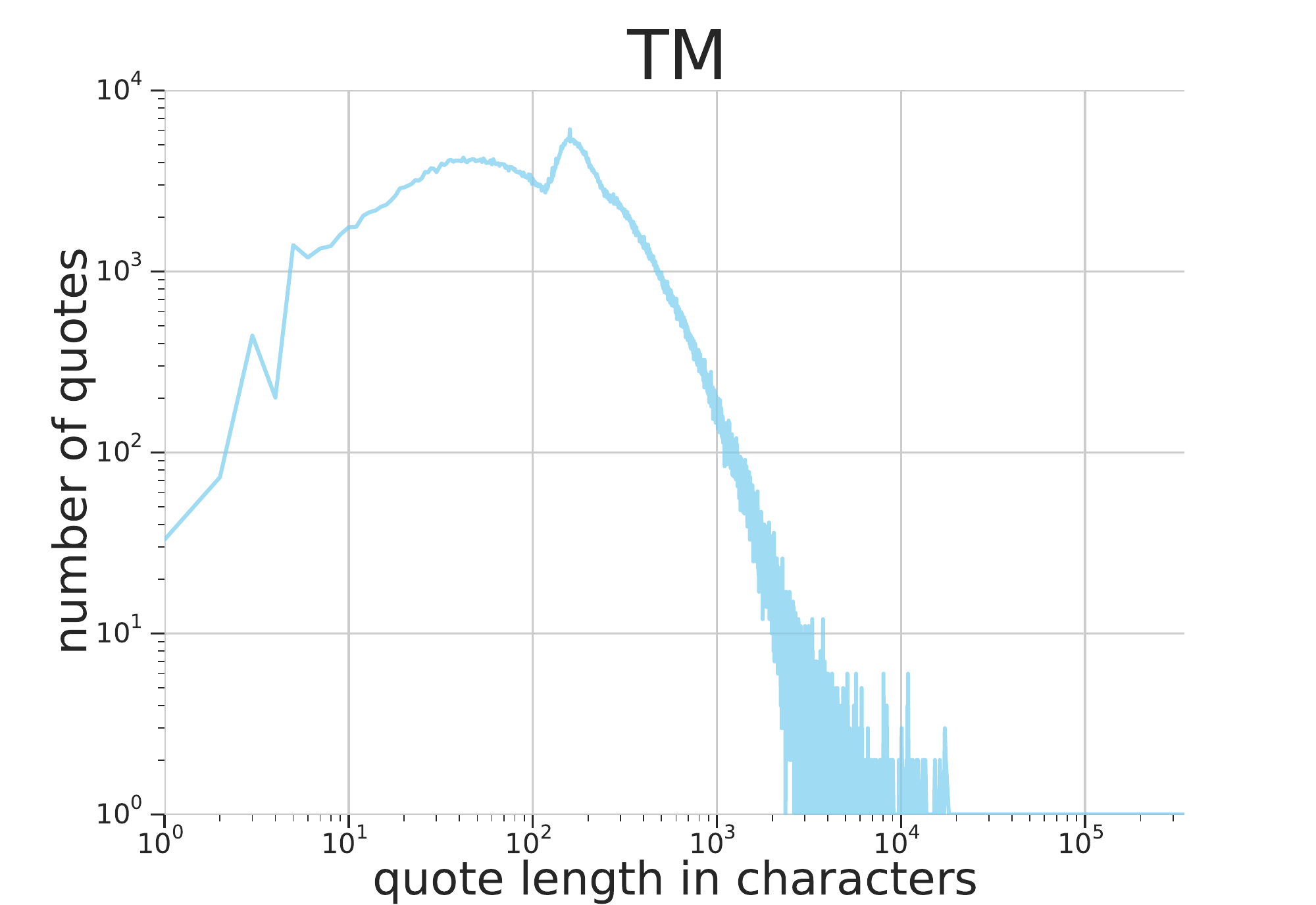}}
\subfloat[]{\includegraphics[width = .5\linewidth]{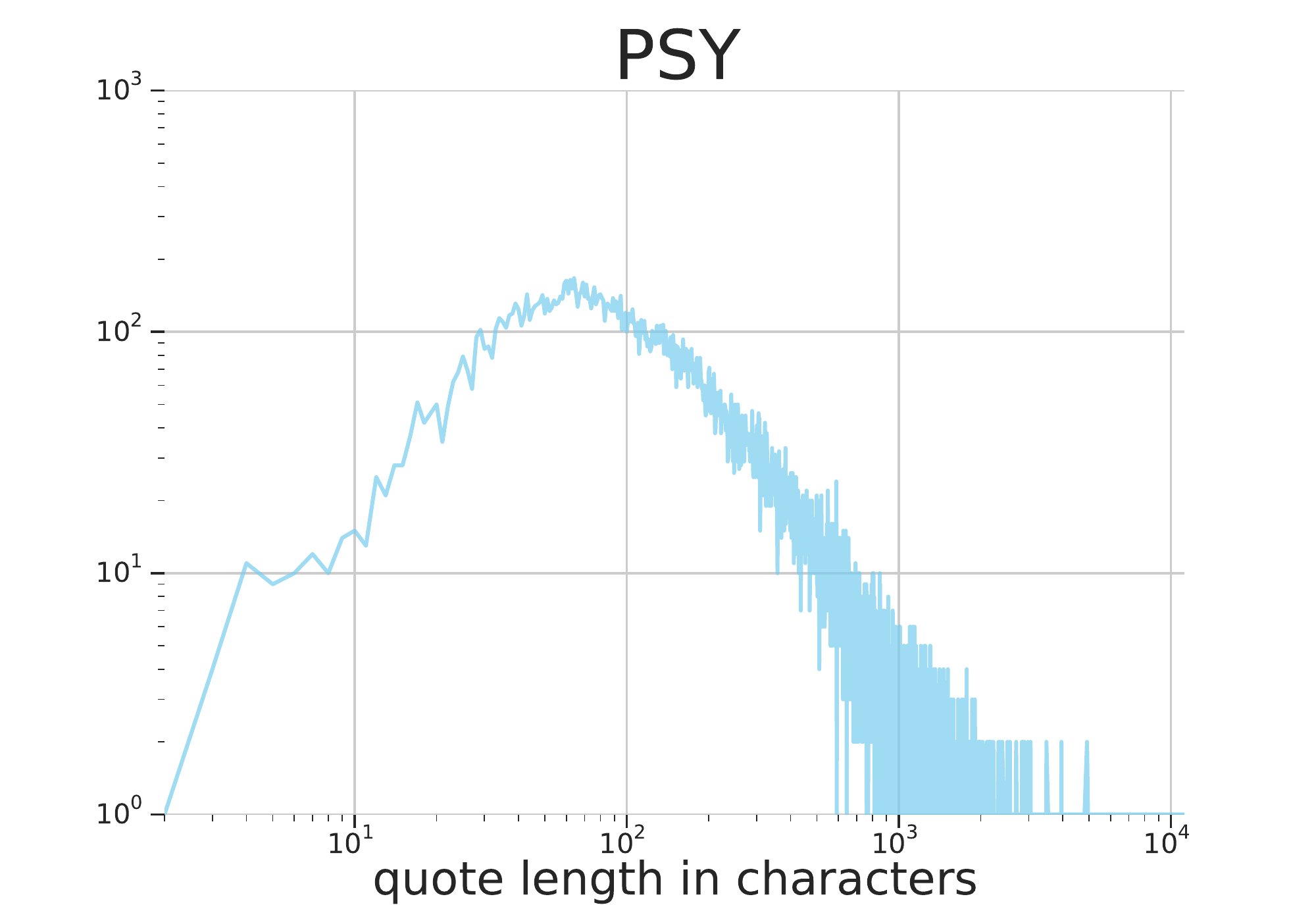}} 
\caption{Quote length distribution (in number of characters)}
\label{fig:quote_length}
\end{figure}

\begin{table}
\centering
\begin{tabular}{rcccc}
 & \textbf{RPG} & \textbf{SWZ}& \textbf{TM}& \textbf{PSY }\\ 
 \scriptsize
\textit{posts} & 14.3M & 1M & 3.6M & 0.15M\\ 
 \scriptsize
\textit{users} & 56.9K & 29.9K & 14.9K & 2.8K \\ 
 \scriptsize
\textit{threads} & 522.7K & 112.1K & 49.2K & 24K \\ 
 \scriptsize
\textit{quotes} & 8.4M & 21.8K & 1.6M & 31.1K \\ 
 \scriptsize
\textit{timespan (years)} & '00-'13& '02-'14& '01-'14& '04-'14\\
\end{tabular}
\caption{Base for the four datasets. \label{tab:size}}
\end{table}

Obtaining a clean, complete corpus of quotes proved to be a demanding process, and required several iterations of curation of the raw html. Here we summarize the problems encountered, and clarify our data cleaning process.

First, quote format in all four forums changed over time: while at first a quote only included the plain text of the quoted comment, the forums added relatively soon the possibility of referencing the quoted post’s author, and subsequently a link to the quoted post. This was most likely the result of new versions of the vbulletin platform offering a slightly different interface. 

Second, a few quotes featured links to posts that failed to appear in our database. In some cases this was due to the original posts being moved or canceled. In a very few other cases, this was an artifact of crawling threads sequentially, as the quoting posts might have been crawled before the quoted post. 

Third, we found that the plain text in many quotes had been altered, either to shorten the quoted text (e.g. replacing text that the quoter deemed irrelevant with ``[snip]'' or ``...''), to correct or emphasize some portion of the original post, or for other reasons. 

Fourth, some users abused the quoting system to reference resources or to provide citations outside of the forum (e.g. to provide a link to another site, or to include an excerpt from a book).

Our data cleaning process proceeded as follows. We discarded nested quotes (quotes embedded within a quote) from the html parse tree. This is a simplifying assumption, yet one that clarifies the interpretability of our results while still adhering to the intuitive definition of quotes. We then extracted the plain text of each quote, the quoted post’s author (when specified), and the link to the quoted post (again when specified). Quotes missing the link to the quoted post were tentatively matched with the latest post in the same thread preceding the quoting post, with a plaintext being superstring of the quoted plaintext, and authored by the author cited in the quote (when specified).

\section{Quote statistics}\label{sec:stats}
\begin{figure}[t]
\subfloat[]{\includegraphics[width = .5\linewidth]{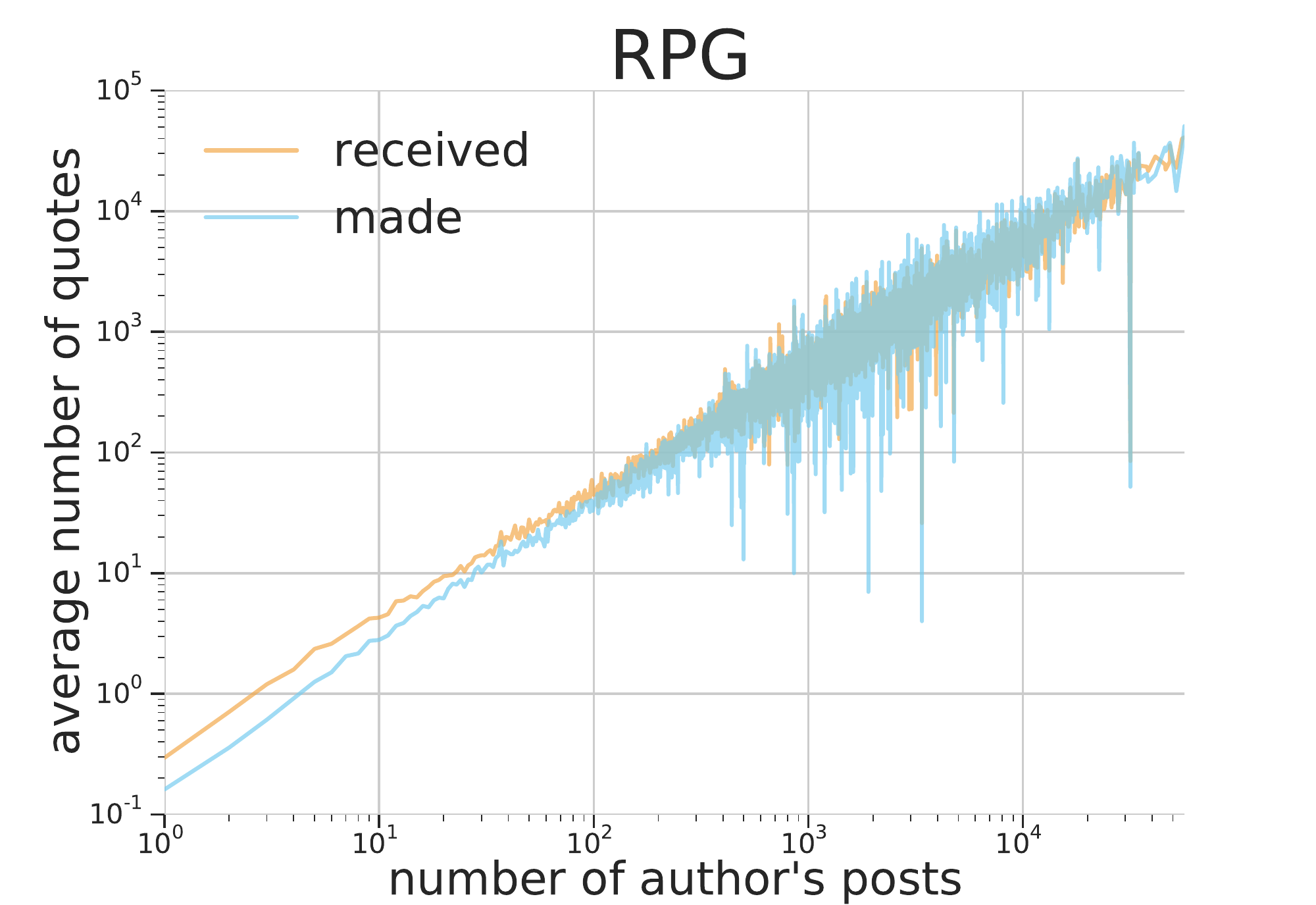}} 
\subfloat[]{\includegraphics[width = .5\linewidth]{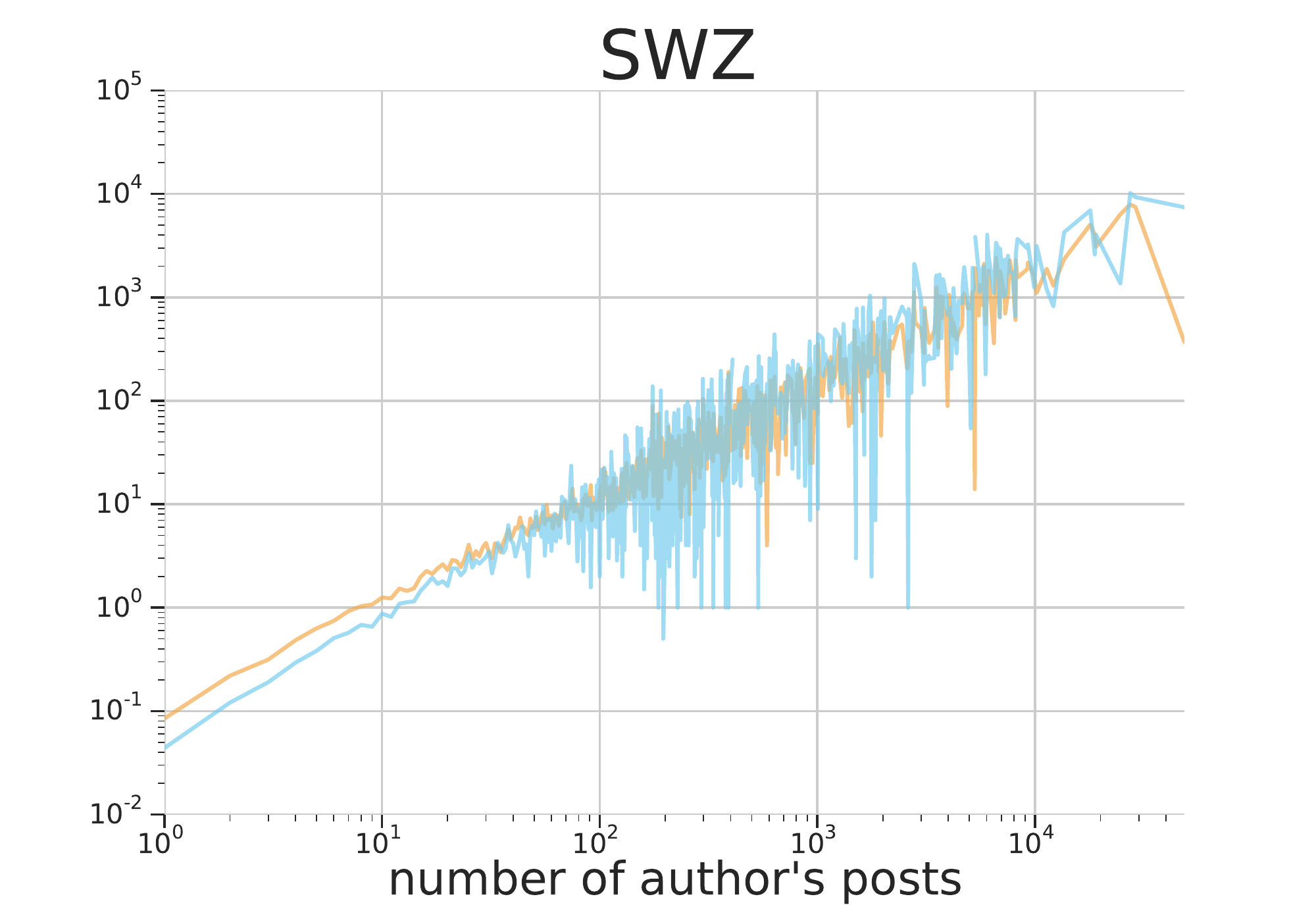}}\    \subfloat[]{\includegraphics[width = .5\linewidth]{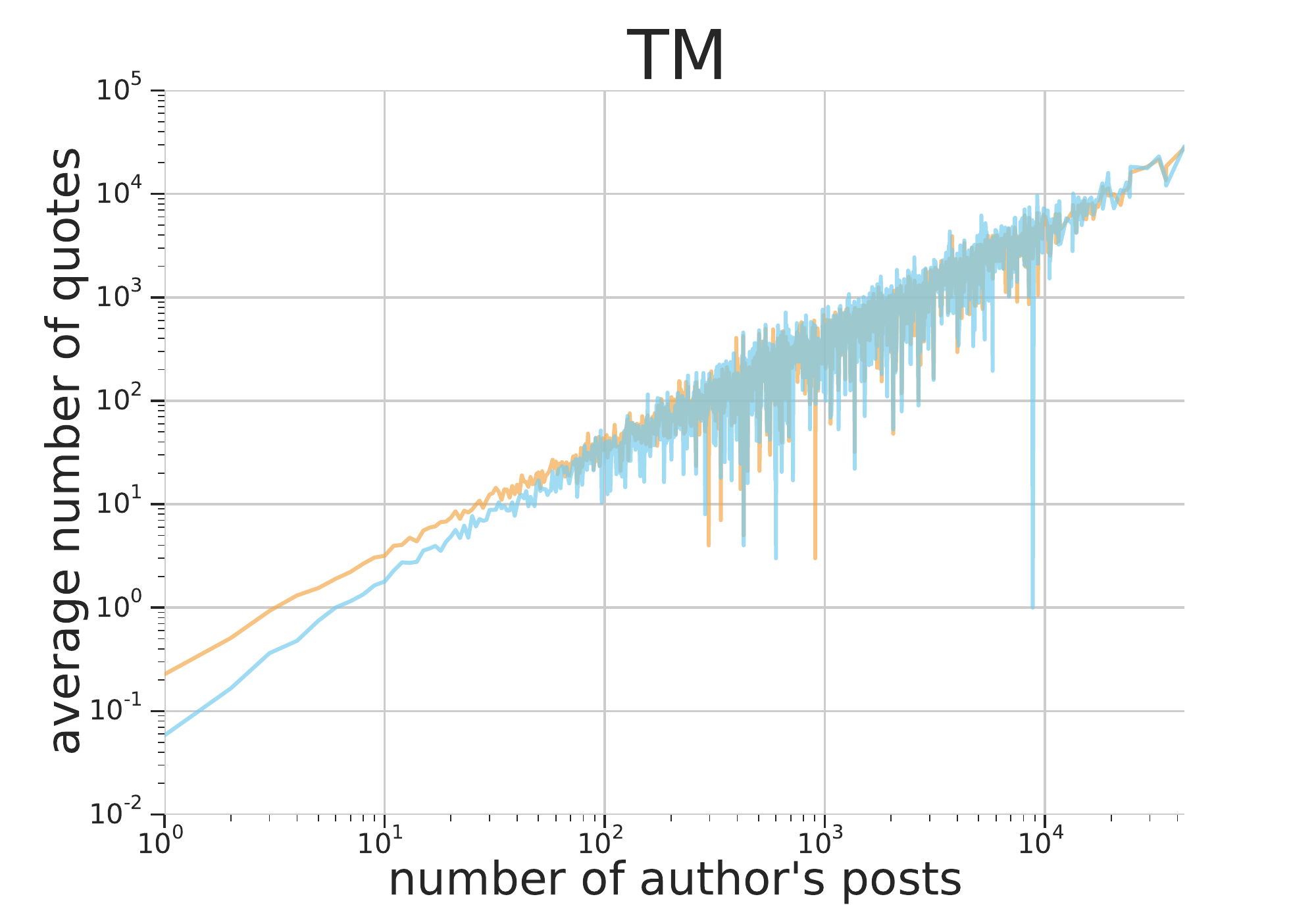}}
\subfloat[]{\includegraphics[width = .5\linewidth]{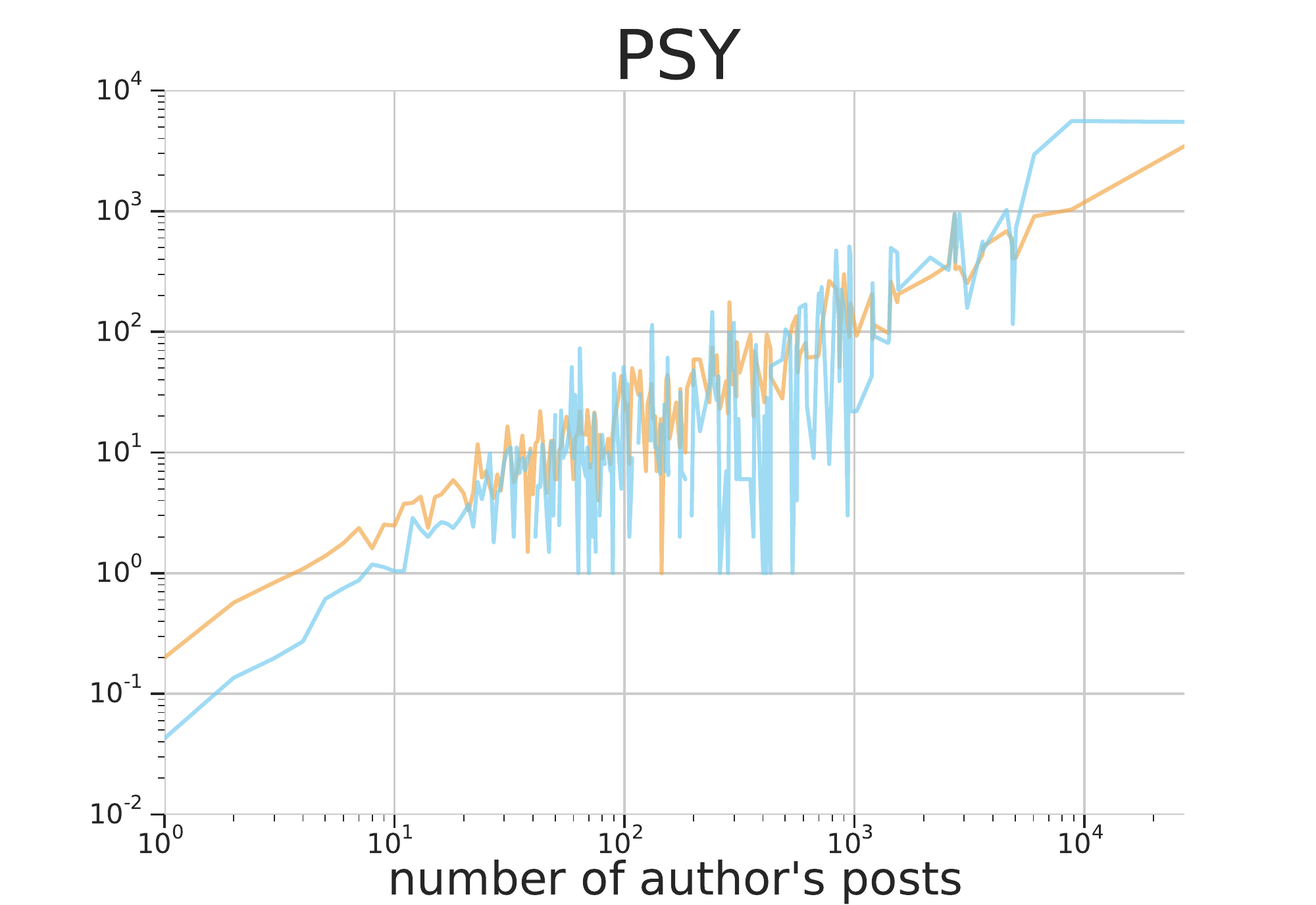}} 
\caption{User number of quotes vs number of posts}
\label{fig:quotes_vs_posts}
\end{figure}

Quote usage varies across different forums, with the ratio of quotes/posts varying between $0.587$ in \rpg and $0.022$ in \swz (see Table \ref{tab:size} and Figure \ref{fig:volume}). The higher quotes/posts ratio of \rpg and \mtl compared to \swz and \psy corresponds to the stronger "hobby chat" character of \rpg and \mtl, as opposed to the more "technical Q/A" character of \swz and \psy. 

\begin{figure}
\subfloat[]{\includegraphics[width = .5\linewidth]{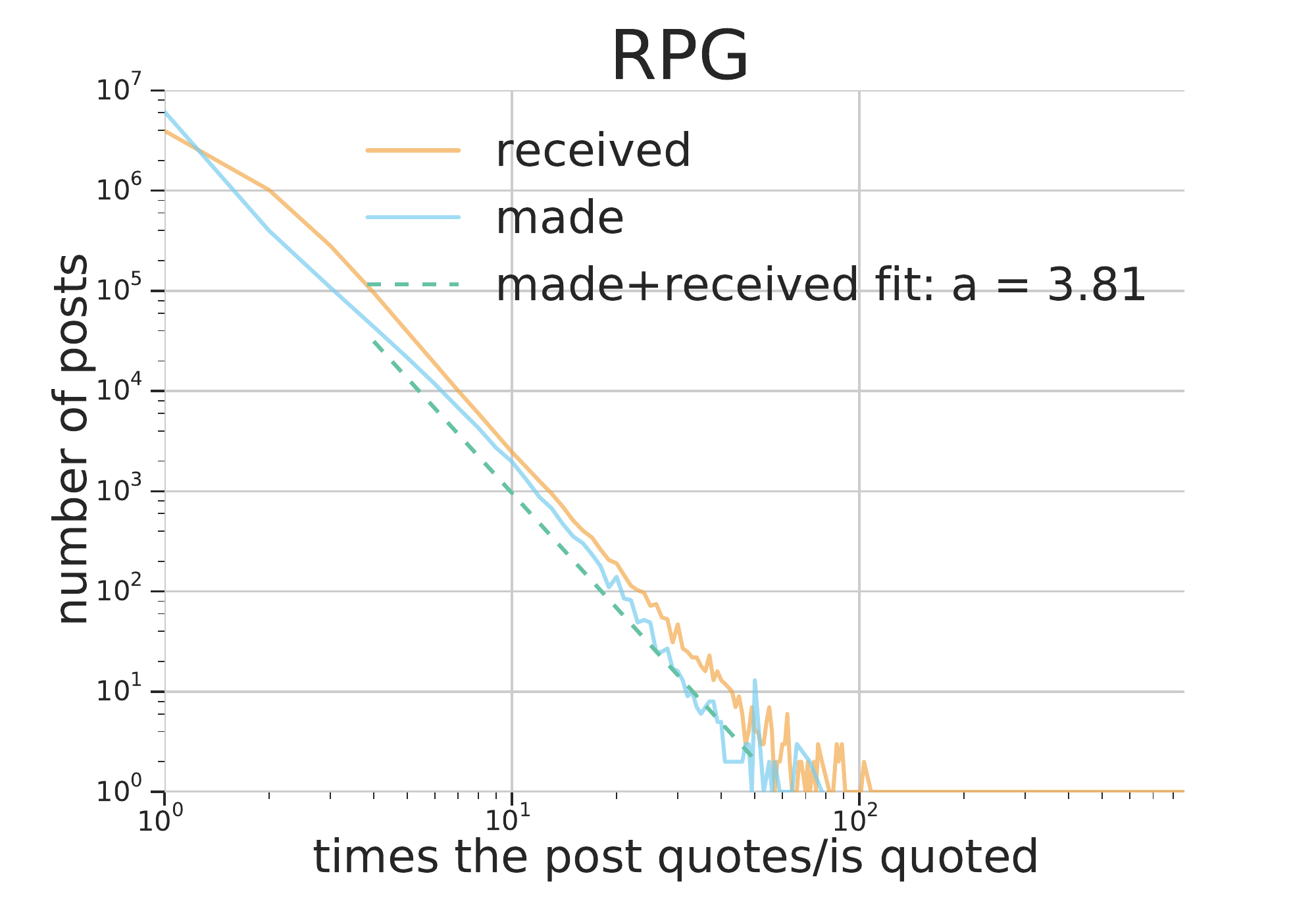}} 
\subfloat[]{\includegraphics[width = .5\linewidth]{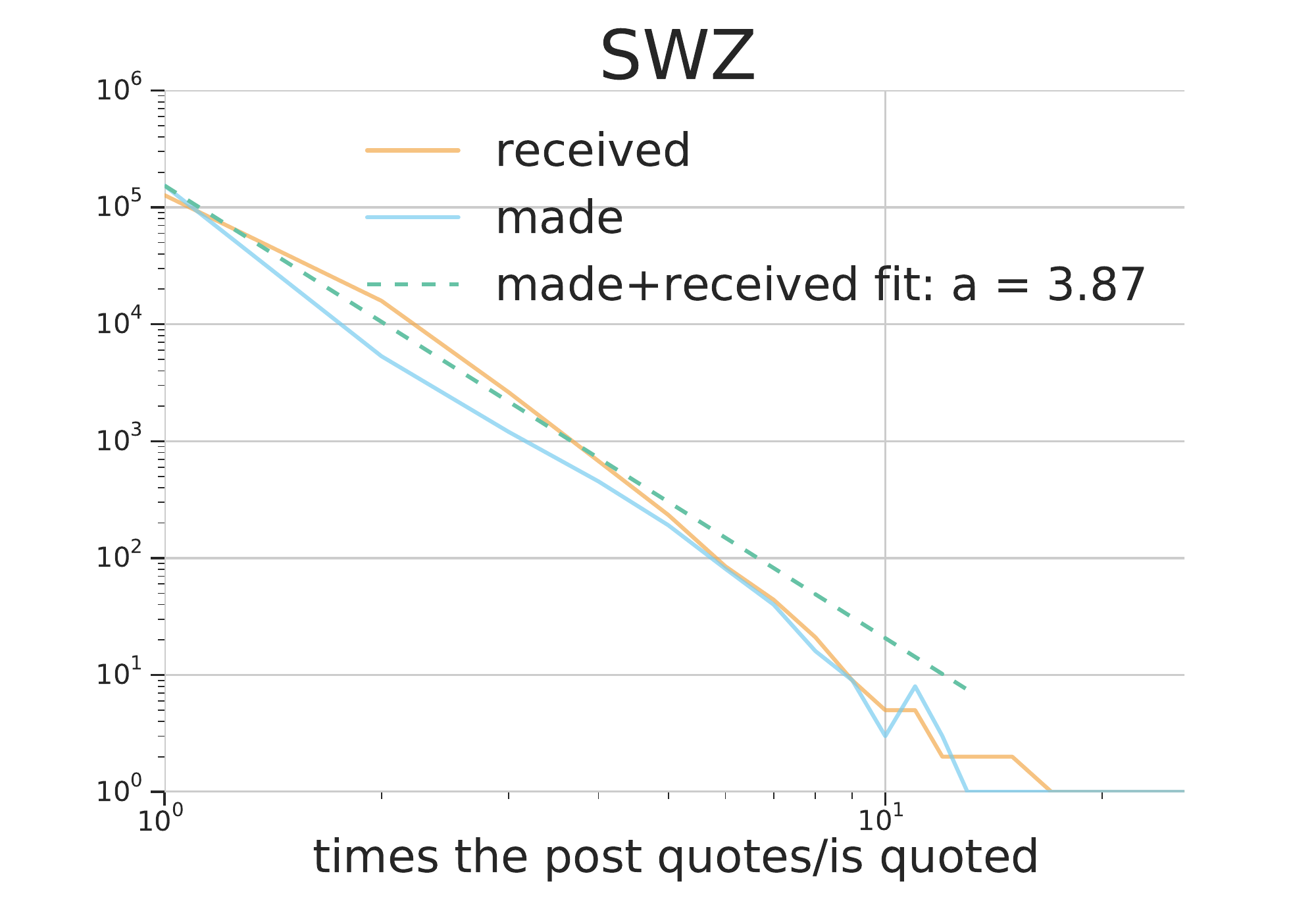}}\    \subfloat[]{\includegraphics[width = .5\linewidth]{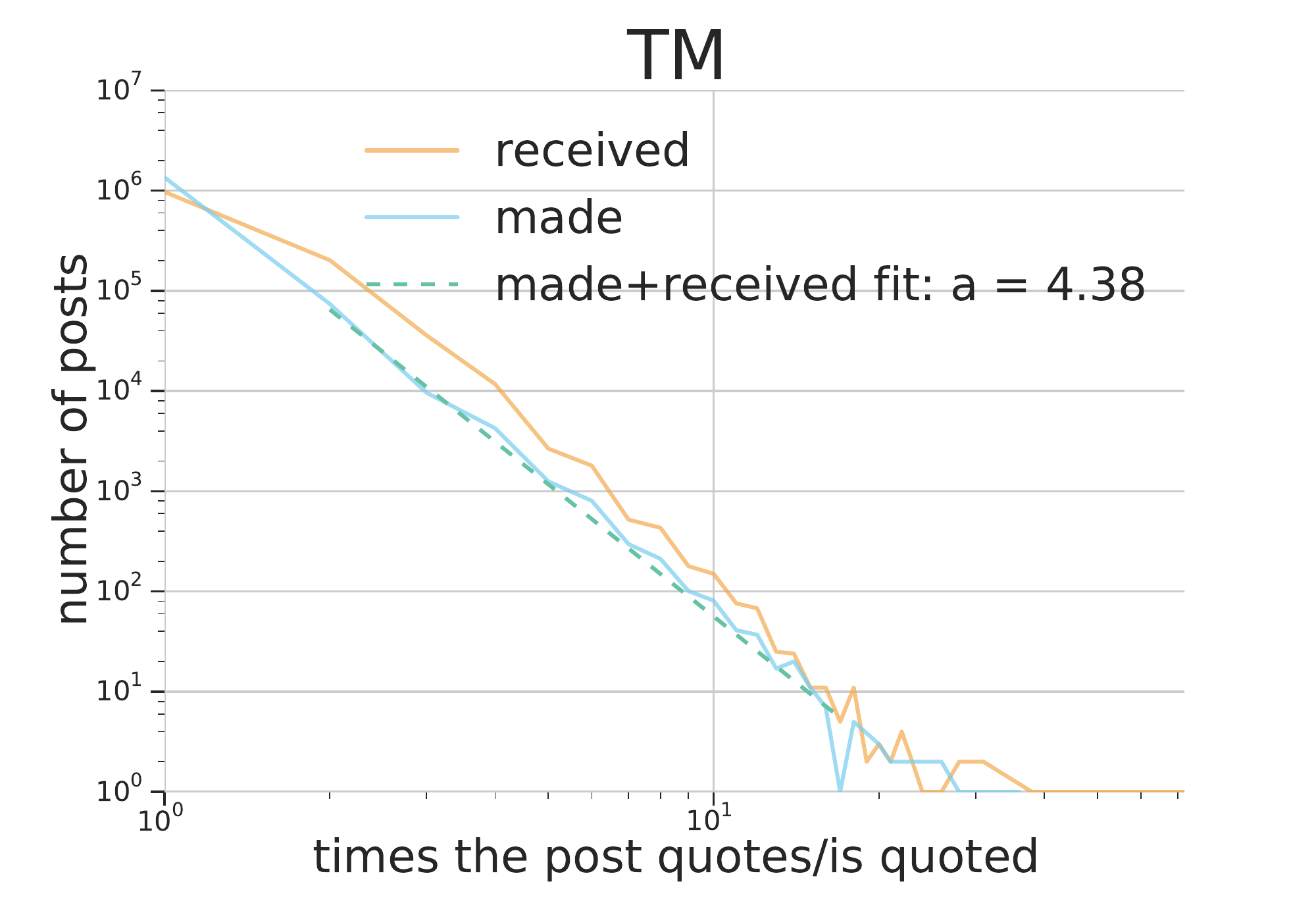}}
\subfloat[]{\includegraphics[width = .5\linewidth]{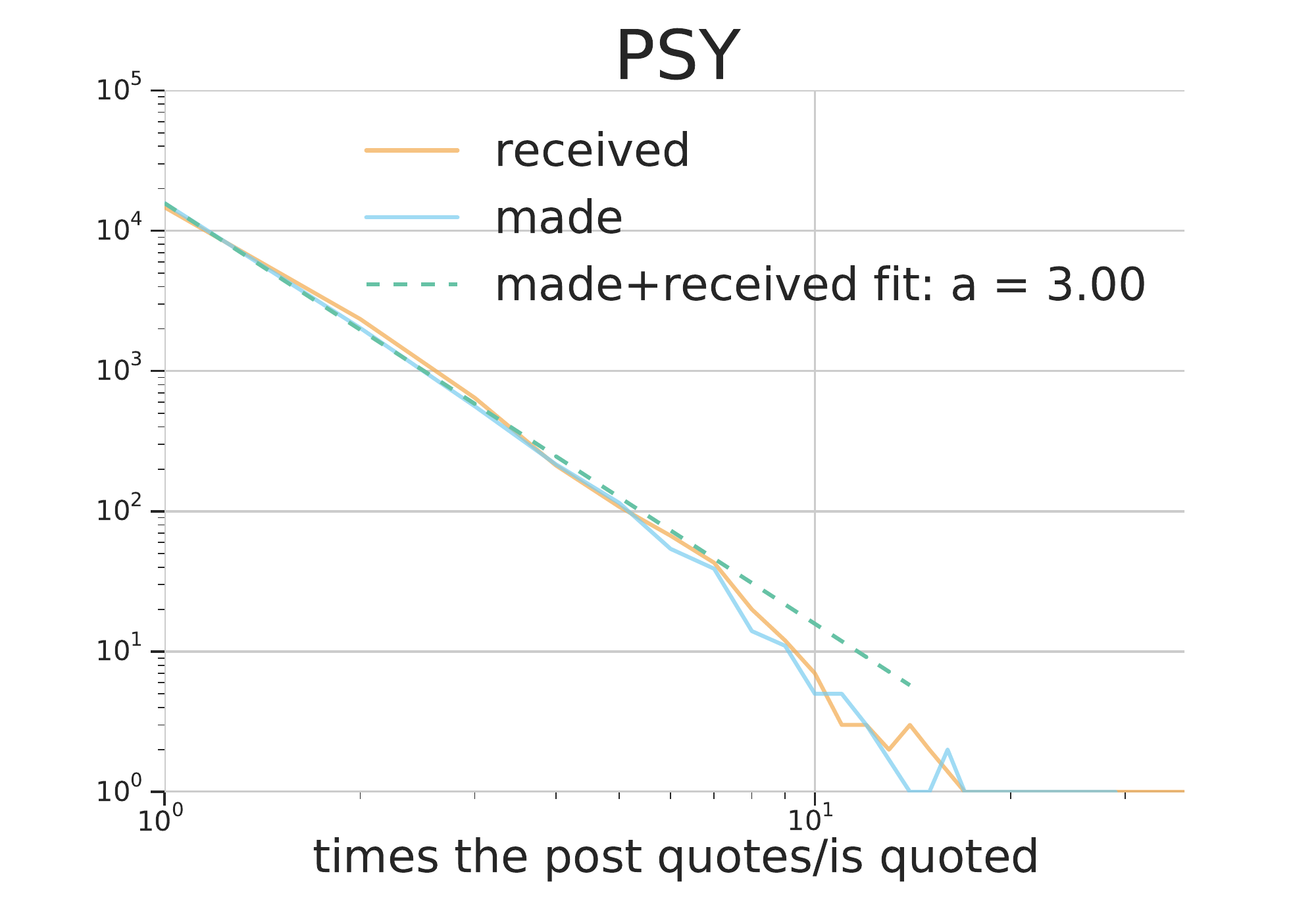}} 
\caption{Quotes per post distribution}
\label{fig:quotes_per_post}
\end{figure}

In all four forums, however, quotes/post appear to follow power-law distributions with exponents ranging between $\approx 3$ (\rpg and \mtl) and $\approx 4$ (\swz and \psy 
-- see Figure~\ref{fig:quotes_per_post}). Interestingly, in each forum the power law exponent for quotes made \emph{to} a post almost perfectly matches that for quotes \emph{by} a post; this is true even at the extreme end of the spectrum, with the exception of a very few highly quoted posts in \rpg (then again, a remarkable post in \rpg \emph{makes} no less than $79$ quotes).
This may be somewhat surprising given that making a quote, as opposed to receiving one, requires some effort by the poster -- and is indeed in contrast with what we observe in many other social contests marked by a similar effort asymmetry, from citation networks to the World Wide Web, where the largest number of citations/links/etc. received by a node typically far outstrips the largest number made.

Another remarkable characteristic of quotes is that \emph{in each forum the ratio of posts/quotes remains eerily constant over time and authors} (see Figures~\ref{fig:volume} and~\ref{fig:quotes_vs_posts}). In particular, more prolific authors receive (and make) more quotes, but no more and no less than groups of less prolific authors with the same total post count -- \emph{there is no ``rich-get-richer'' effect}, again in marked contrast to most other social environments.
This is particularly surprising given that not only does the post count change significantly from month to month, but that the average user "lifetime" (less than $1.5$ years for all four forums) is significantly shorter than the time interval under observation. Quote/post ratio then appears to be an extremely specific signature of each forum's language and interaction patterns, suggesting the existence of an independent "geist" of each forum that, although emerging from the behaviour of individual posters, assumes and actively maintains a relatively unchanging identity of its own by shaping the behaviour of subsequent generations of posters.

\begin{figure}
\subfloat[]{\includegraphics[width = .5\linewidth]{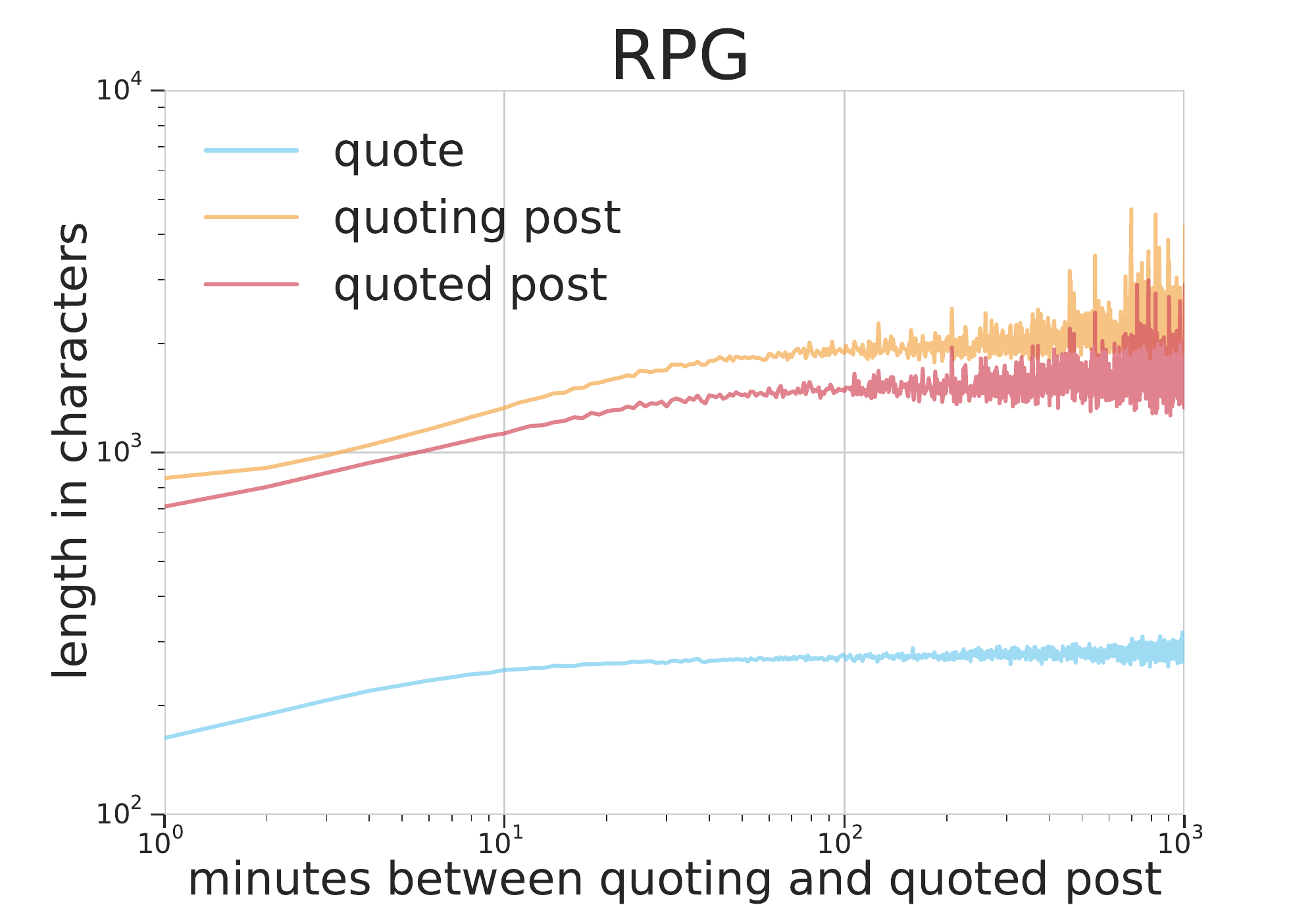}} 
\subfloat[]{\includegraphics[width = .5\linewidth]{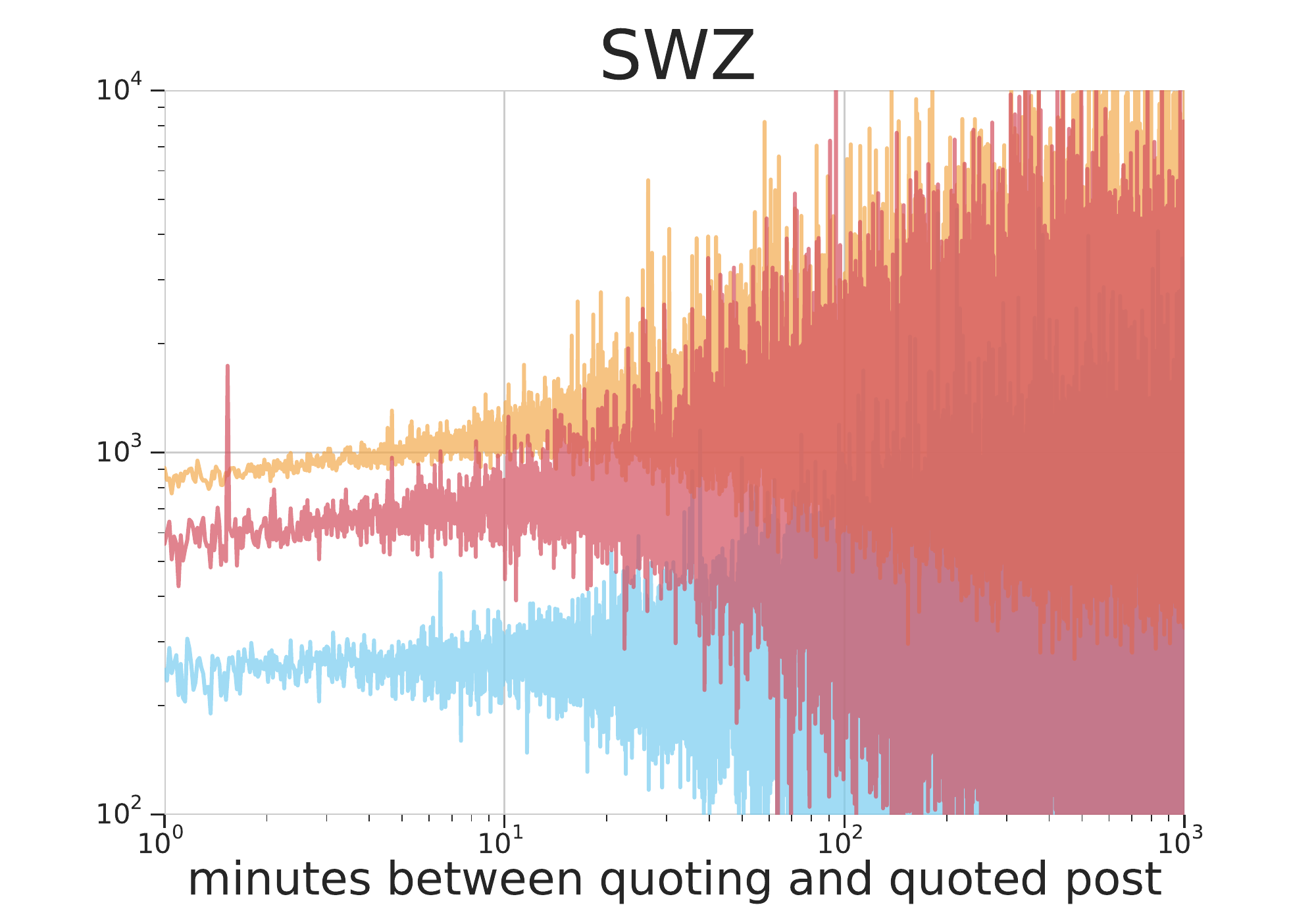}}\    \subfloat[]{\includegraphics[width = .5\linewidth]{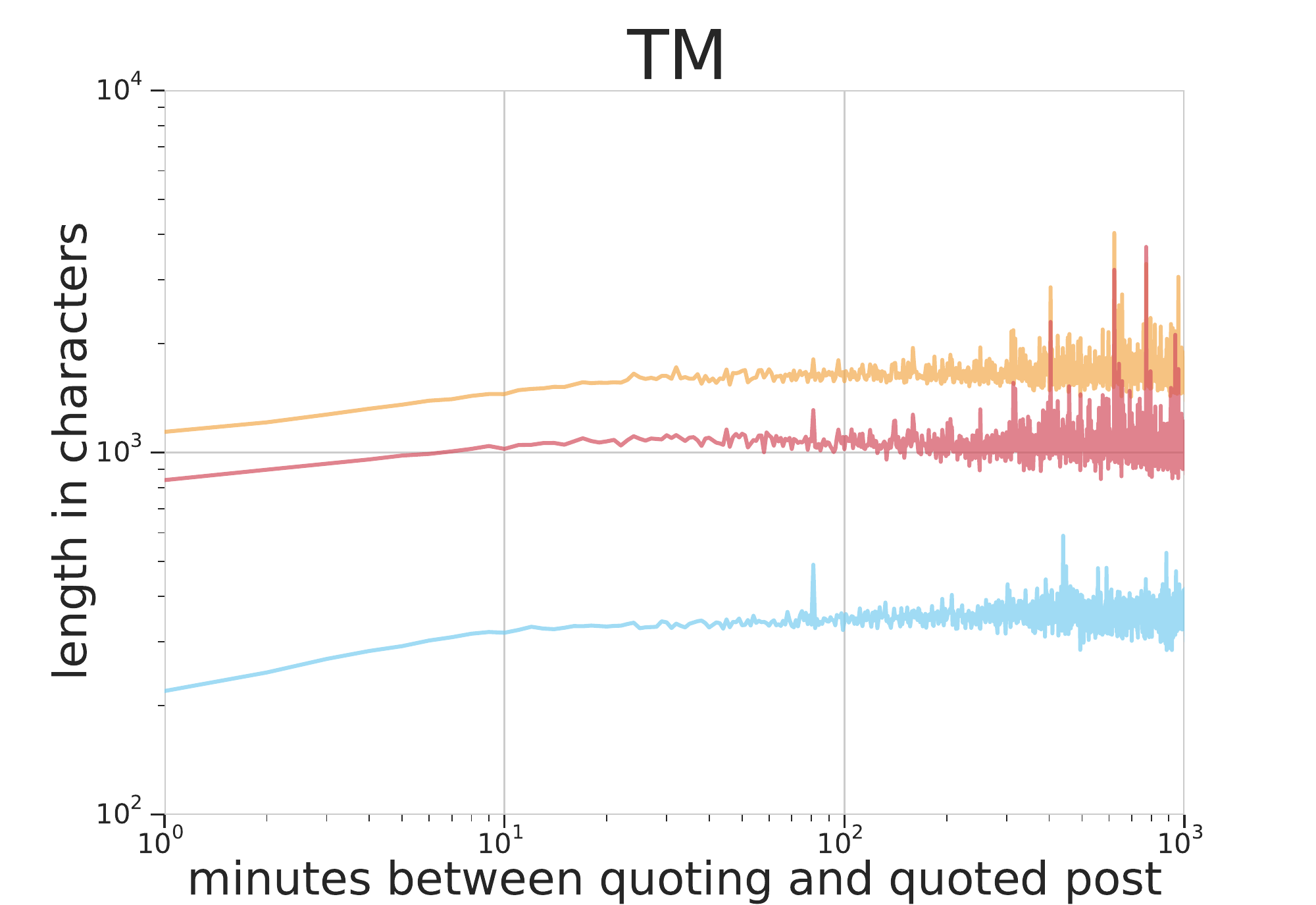}}
\subfloat[]{\includegraphics[width = .5\linewidth]{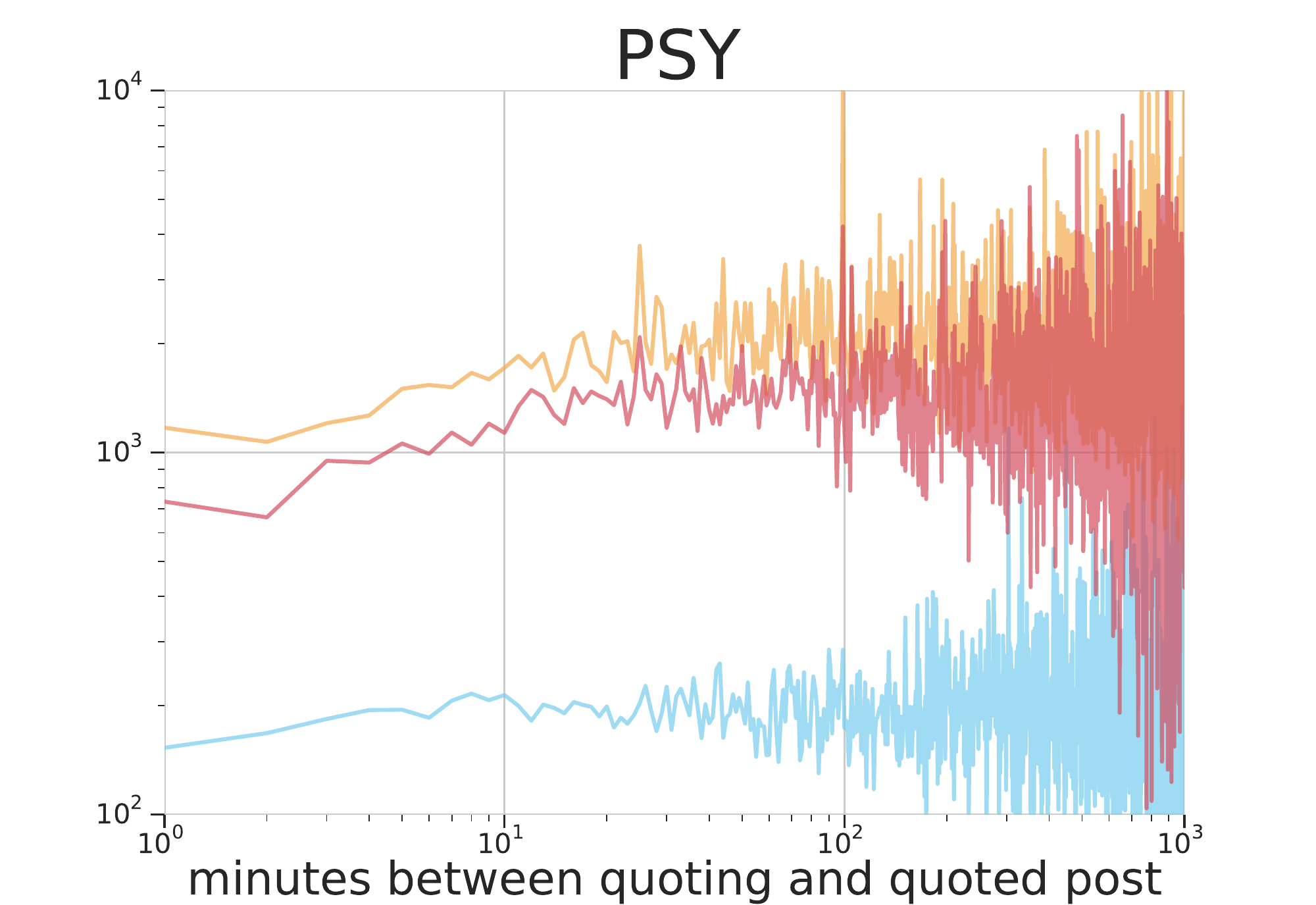}} 
\caption{Quote time difference vs quoting, quoted post, and quote length}
\label{fig:time_difference}
\end{figure}
 
\begin{figure}
\subfloat[]{\includegraphics[width = .5\linewidth]{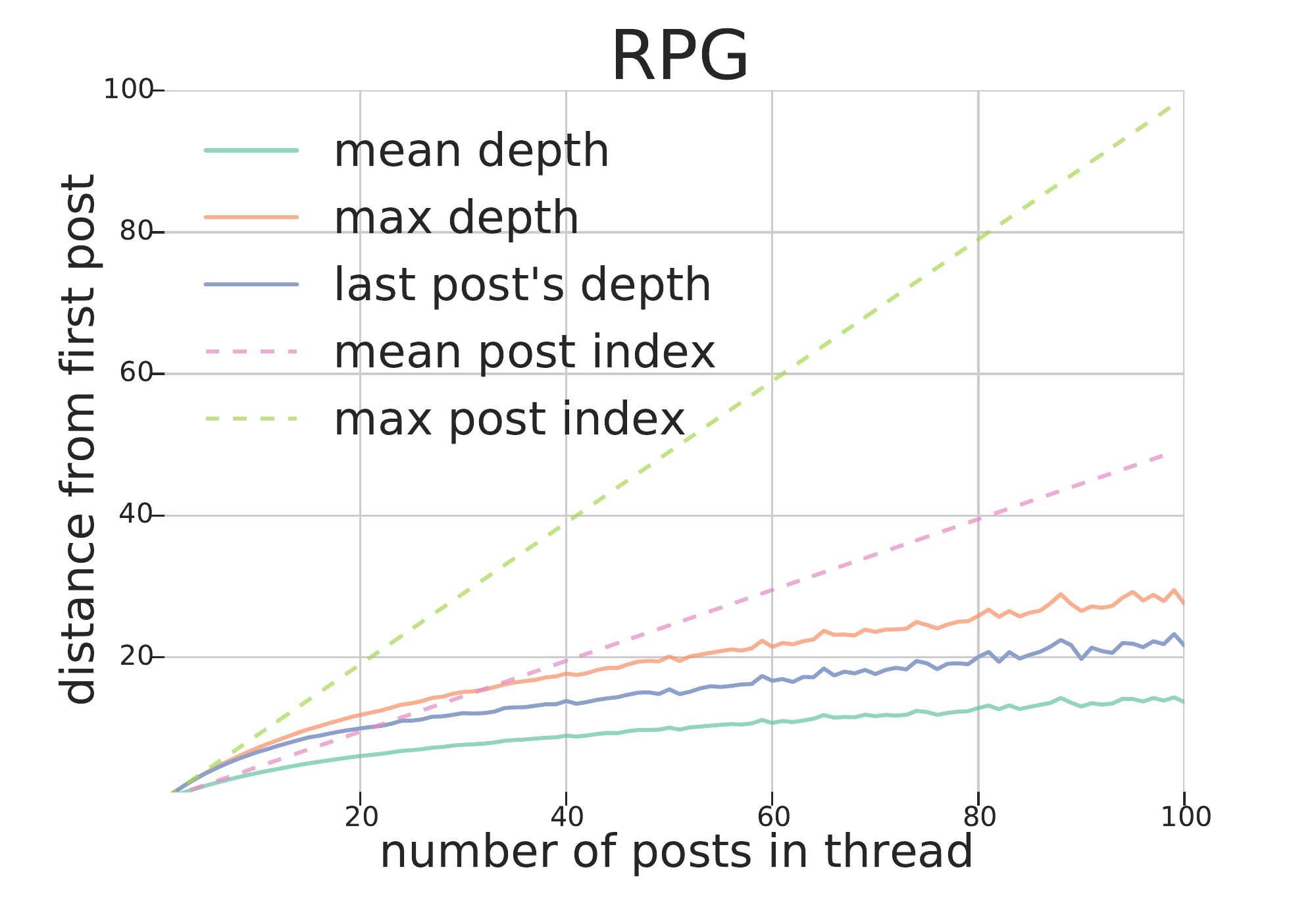}} 
\subfloat[]{\includegraphics[width = .5\linewidth]{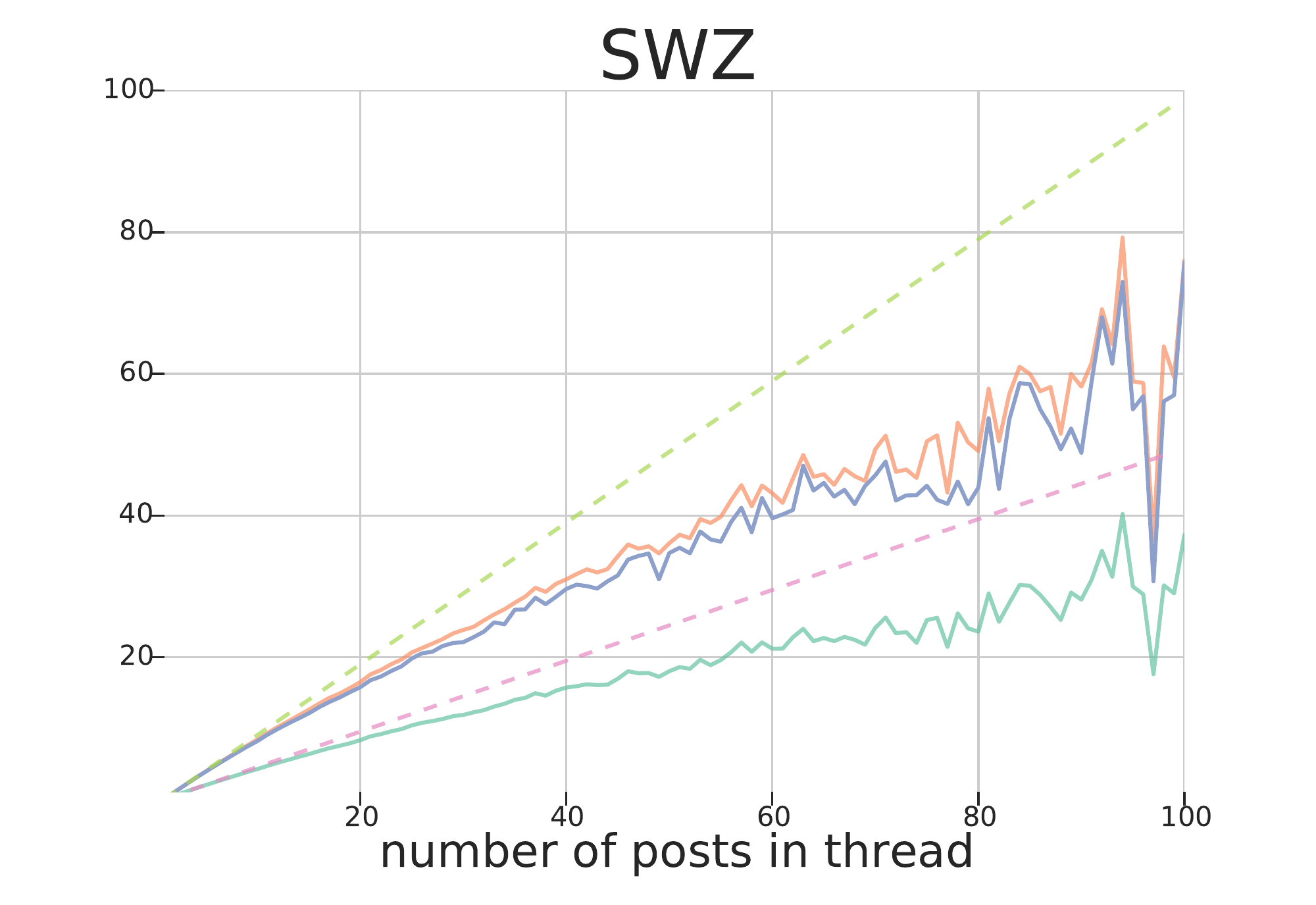}}\    \subfloat[]{\includegraphics[width = .5\linewidth]{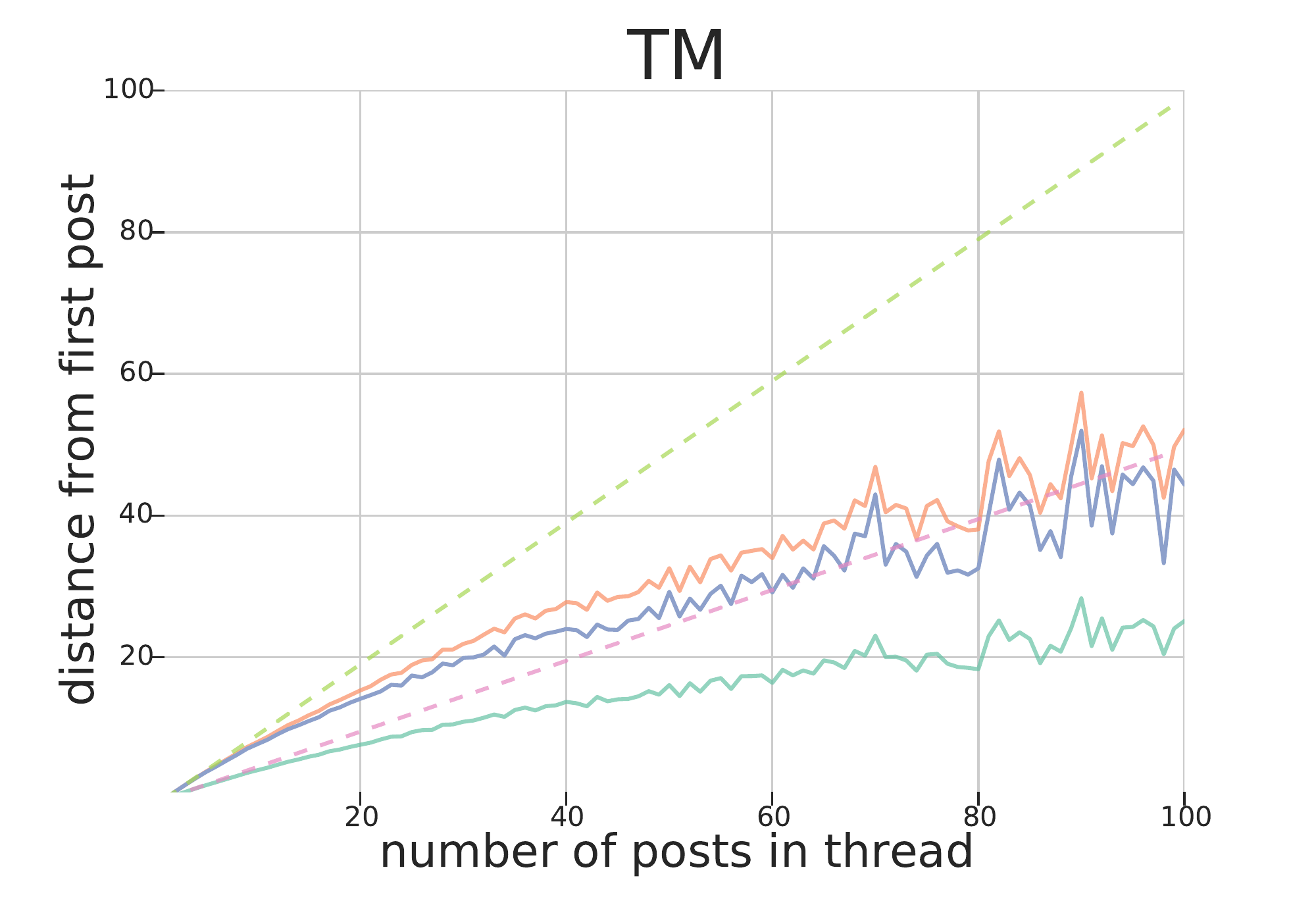}}
\subfloat[]{\includegraphics[width = .5\linewidth]{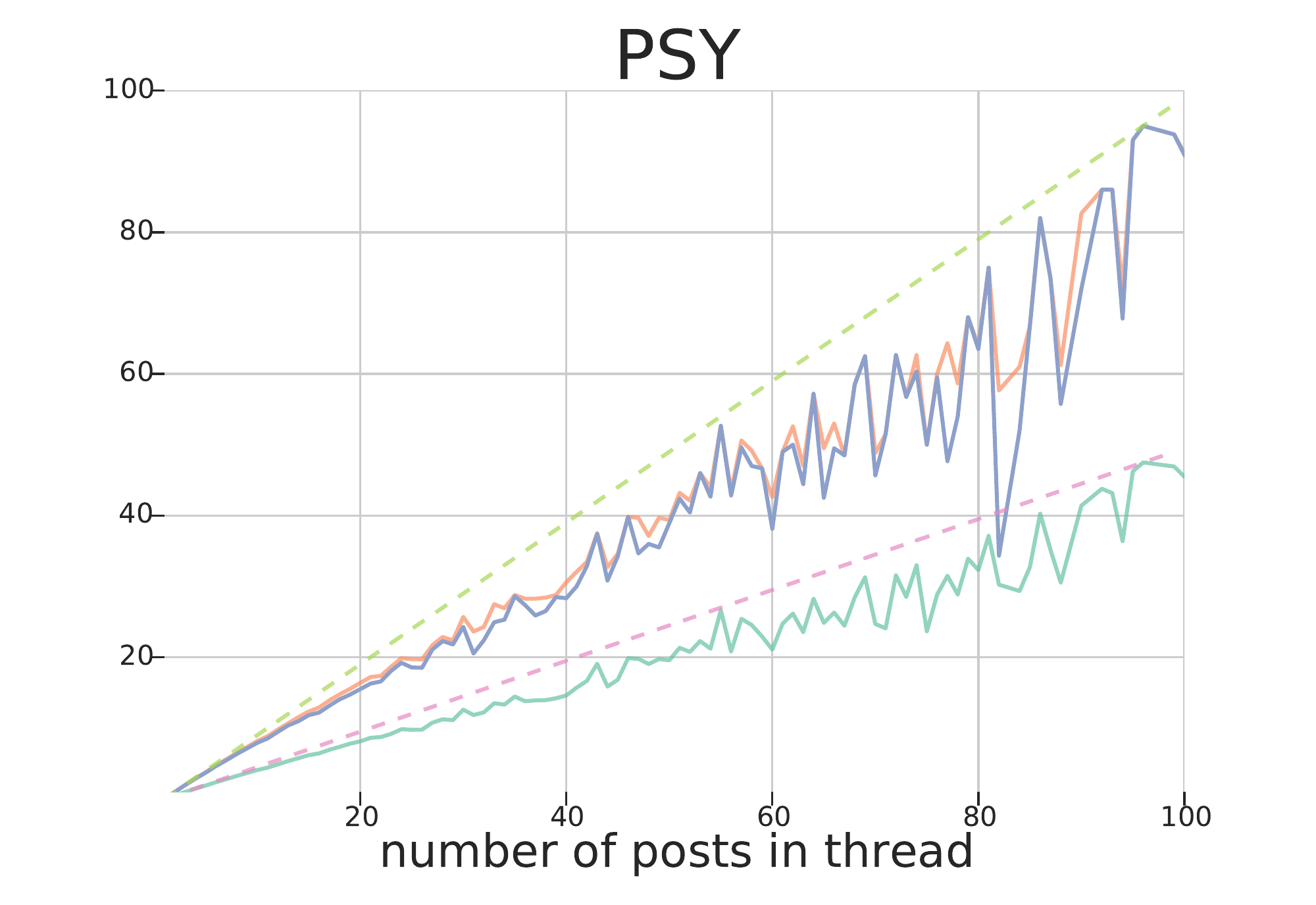}} 
\caption{Thread depth in maximum, average, and last post's quote distance from the first post}
\label{fig:depth}
\end{figure}




\section{Navigation}\label{sec:nav}

\begin{figure}
\subfloat[]{\includegraphics[width = .5\linewidth]{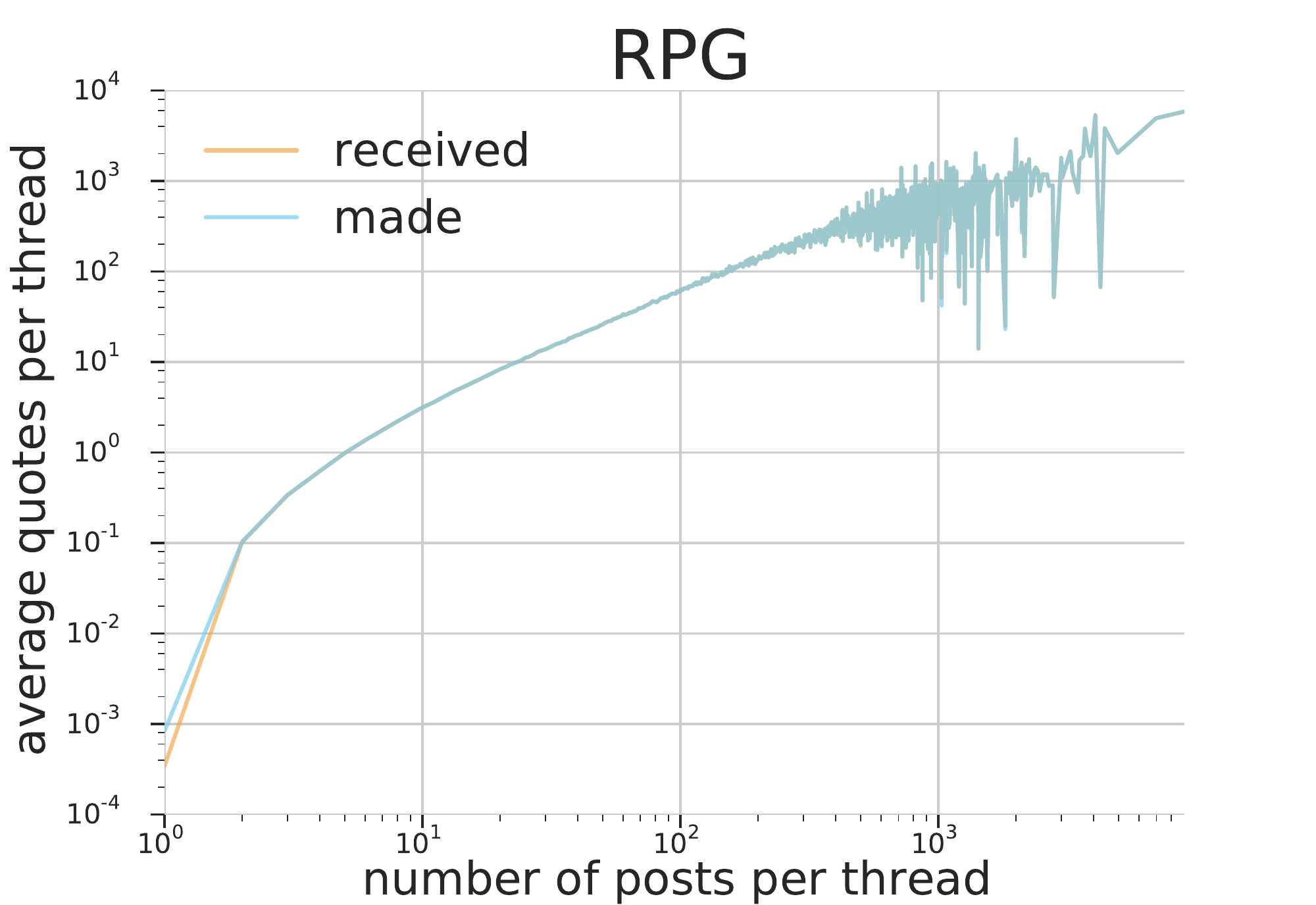}} 
\subfloat[]{\includegraphics[width = .5\linewidth]{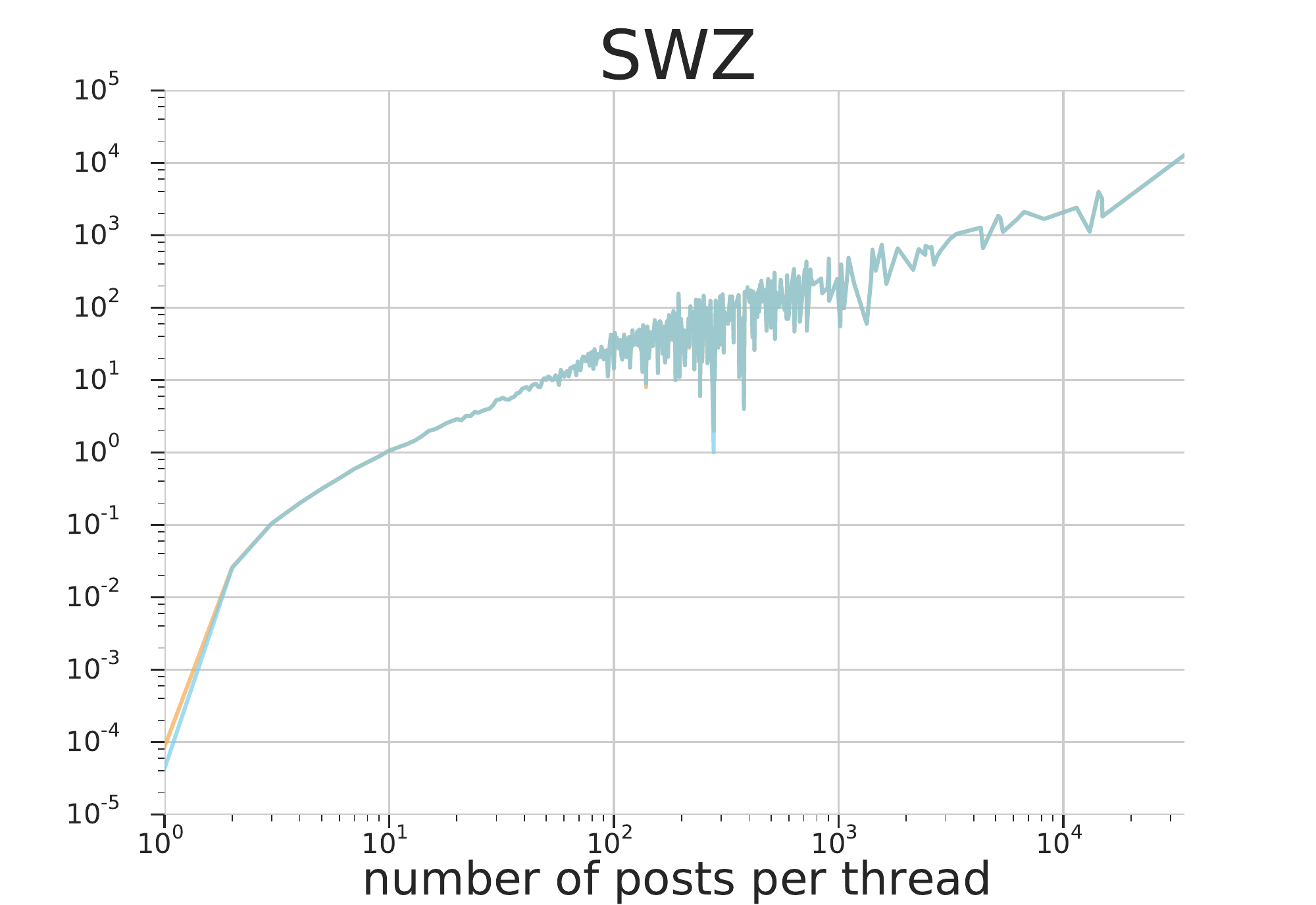}}\    \subfloat[]{\includegraphics[width = .5\linewidth]{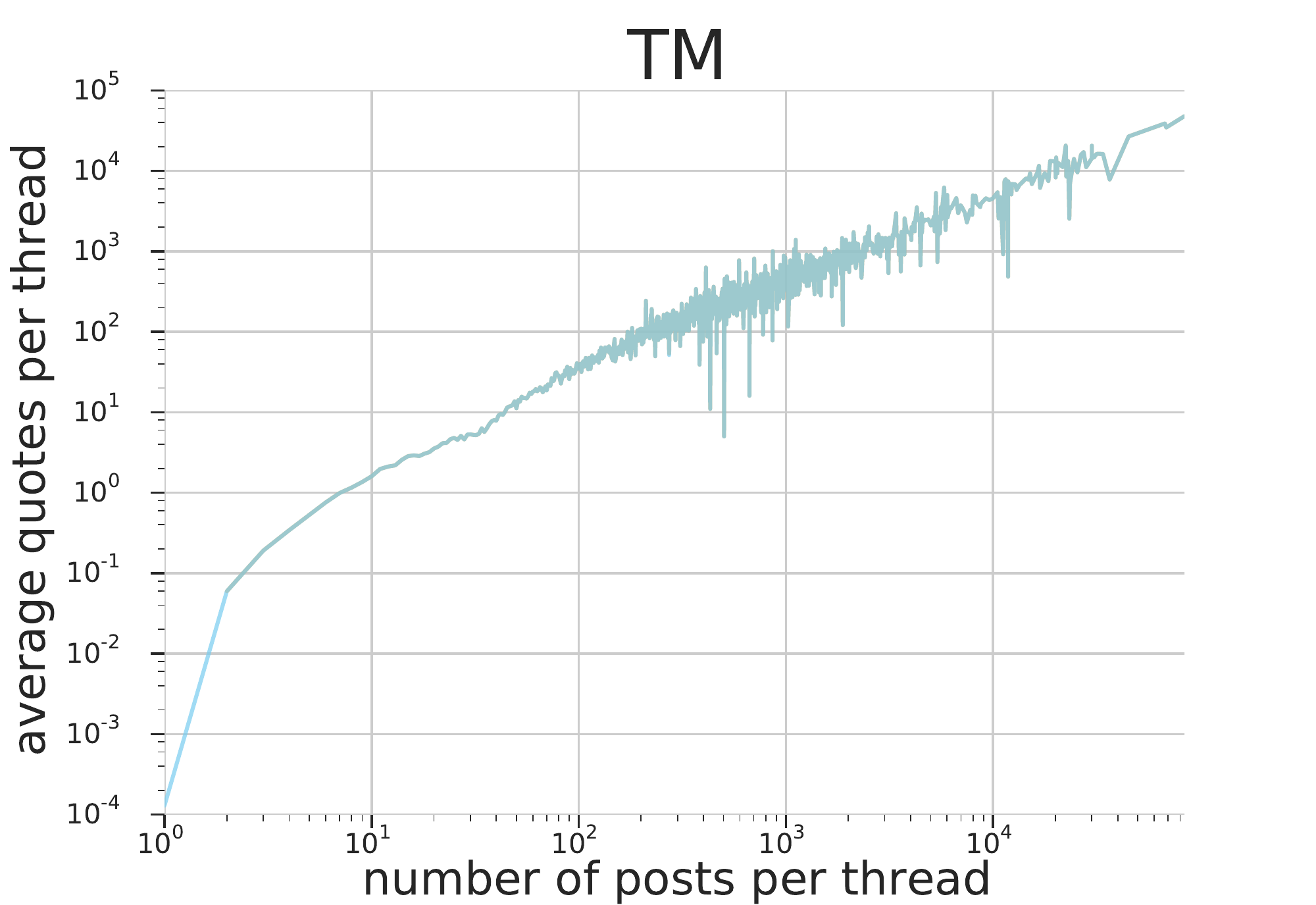}}
\subfloat[]{\includegraphics[width = .5\linewidth]{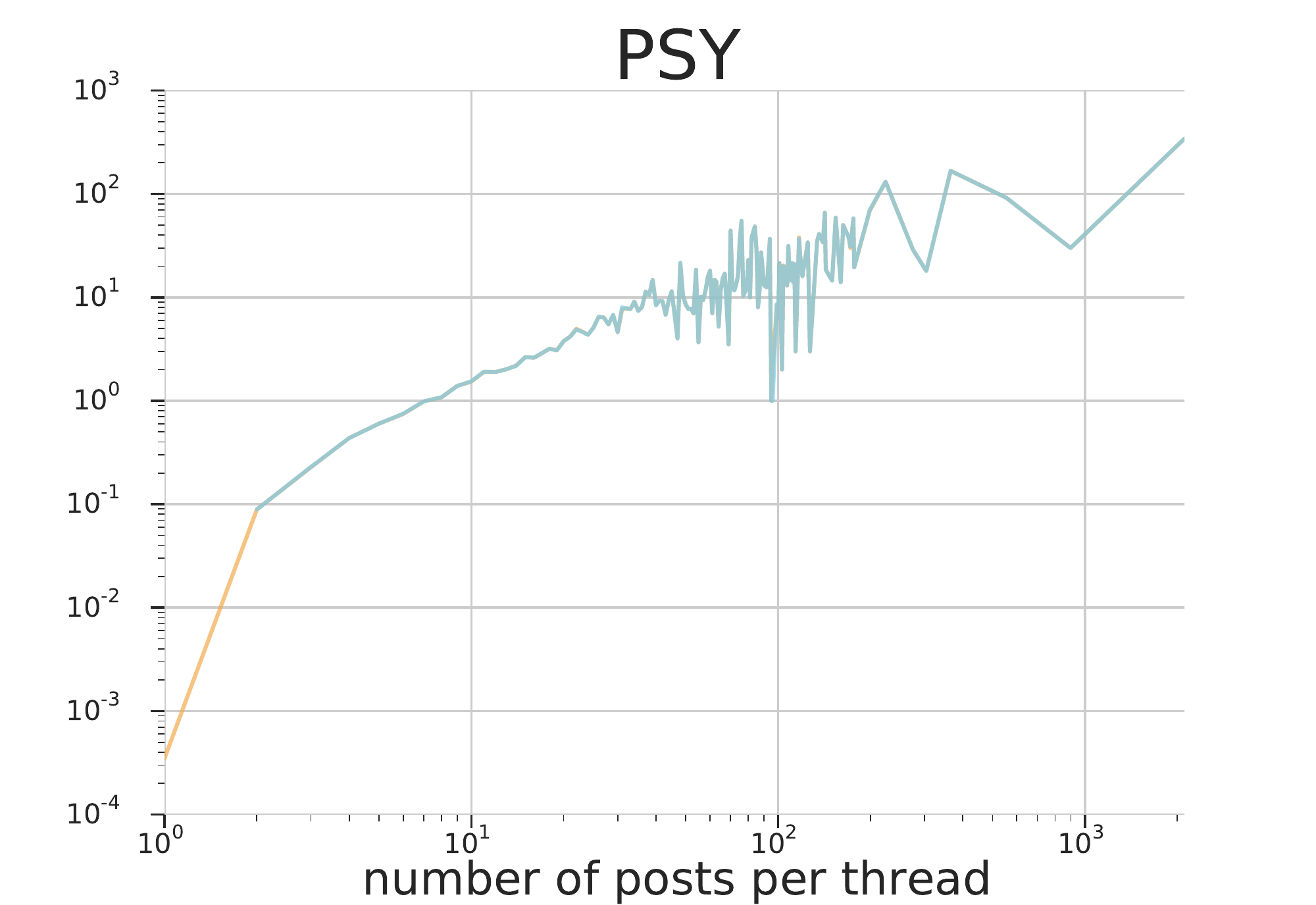}} 
\caption{Average number of quotes per number of posts per thread}
\label{fig:quotes_vs_thread_length}
\end{figure}

Even though quotes by/to an author precisely track that author's post count, the same cannot be said of quotes by/to a thread.  Short threads both make and receive relatively fewer quotes per post (see Figure~\ref{fig:quotes_vs_thread_length}). A possible explanation is that \emph{a unique role of quotes is to aid intra-thread navigation} -- with shorter threads being intrinsically easier to navigate and thus requiring less quote support.

Further analysis of quote length supports this hypothesis. Quote length follows a power-law distribution, at least beyond a minimum threshold of a $140$ characters\footnote{using the python module "powerlaw": arXiv:1305.0215} (see Figure~\ref{fig:quote_length}) -- shorter quotes are comparatively rarer, showing the difficulty of conveying meaningful information with a chunk of text shorter than a tweet. While a few of the very shortest quotes are essentially typing/posting errors, the majority of quotes of even $2$ characters appear valid (e.g. ``no'', ``3?'', ``me''); most of these tiny quotes refer to a very ``close'' post on which they rely to provide the appropriate context. And indeed, quote length markedly grows with the temporal distance between quoting and quoted post (see Figure~\ref{fig:time_difference}).

Another way to observe this phenomenon is to consider the \emph{depth} of posts, defined for the initial post of any thread as $0$, and for any other post $p$ as $1$ plus the minimum depth of any post that $p$ quotes or immediately follows in the thread -- in some sense, the depth of a post being the length of the shortest discussion leading to that post. Without quotes, both maximal and average post depth would be proportional to thread length. However, in practice, quotes provide shortcuts in the discussion, significantly shortening longer threads more than short ones, both in average and maximal post depth (see Figure~\ref{fig:depth}). It is not entirely clear whether (forums with) longer threads tend to generate more quotes, or instead (forums whose culture generates) abundant quotes can more easily sustain longer threads -- but it seems evident from~Figure~\ref{fig:depth} that in forums where threads are on average longer (like \textit{RPG} and \textit{Truemetal}), the overall number of quotes is comparatively higher and thread depth is consistently kept small; conversely, in forums with a smaller average number of posts per thread, like \textit{Swzone} and \textit{Psychlinks}, this effect is less prominent.

\section{Social structure}\label{sec:social}
For many years, forums have been the venue of choice for communities of users sharing interests on a topic. However, forum users can interact almost only through discussion -- forums mostly lack the trappings of modern “social” platforms, such as friendship, liking, and reputation mechanisms. Even if quotes are indicators of attention, common interest, and attribution, forums do not tally them: there is no immediate way to learn which or even how many users have quoted a given user or post. In this light, it may be surprising that it is possible to retrieve a latent structure of a forum’s community by observing how users quote one another, and that this structure shows the typical features of a social network, as we shall see in the next subsection.

\begin{table}[htbp]
\centering
\begin{tabular}[bp]{l c c c c}

 				   & \textbf{RPG}  & \textbf{SWZ}  & \textbf{TM}   & \textbf{PSY}  \\
\scriptsize
 Nodes                      & 35118      & 11544    & 9661     & 1553     \\
\scriptsize
 Edges                      & 2.5M    & 50.8K    & 291.7K   & 5983     \\
\scriptsize
 Zero InDeg Nodes           & 3330       & 1832     & 533      & 117      \\
\scriptsize
 Zero OutDeg Nodes          & 9628       & 6084     & 2996     & 853      \\
\scriptsize
 NonZero Deg Nodes   & 22.2K      & 3628     & 6132     & 583      \\
\scriptsize
 Unique directed edges      & 2.5M    & 50.8K    & 291.7K   & 5983     \\
\scriptsize
 Unique undirected edges    & 1.8M    & 41.5K    & 203.9K   & 4804     \\
\scriptsize
 Self Edges                 & 4174       & 766      & 1265     & 70       \\
\scriptsize
 BiDir Edges                & 1.4M    & 19.3K    & 176.8K   & 2.4K     \\
\scriptsize 
 Closed triangles           & 176.8M  & 250K   & 5.4M  & 11.4K    \\
\scriptsize
 Open triangles             & 1.4G & 11.9M & 51.2M & 715K   \\
\scriptsize
 Frac. of closed triads     & 0.111   & 0.021 & 0.0962 & 0.0158 \\
\scriptsize
 Conn. comp. size:  & 0.995   & 0.975 & 0.993 & 0.993 \\
\scriptsize
 Strong conn. comp. size:   & 0.625   & 0.283 & 0.625 & 0.365 \\
\scriptsize
 Approx. full diameter:     & 7          & 7        & 7        & 6        \\
\scriptsize
 90\% effective diameter:    & 3.317   & 3.690 & 3.338 & 2.922 \\
 \scriptsize
 Average clustering:    & 0.385   & 0.305 & 0.469 & 0.431 \\
 \scriptsize
 Assortative mixing:	&0.088&-0.249&0.226&-0.004\\
\end{tabular}
\caption{Quote network statistics}
\label{tab:quote_net}
\end{table}

\subsection{The author-quote graph}
We define the \textit{author-quote graph} as the directed, weighted graph that has users as nodes, has an edge between user $a$ and user $b$ if $a$ has ever quoted one of $b$'s posts, with weight equal to the total number of times $a$ has quoted $b$.

As Table \ref{tab:quote_net} shows, the quote networks obtained from the four forums sport many characteristics of social networks. 

First, the graphs are sparse, containing only a small fraction of all potential edges. Second, they are small worlds, with a giant connected component. More precisely, all forums show a weakly connected component  that includes more than $97\%$ of all nodes. The strongly connected components include approximately $62\%$ of all nodes in the case of \textit{RPG }and \textit{Truemetal}. These numbers closely match the corresponding values, $92\%$ and $68\%$ respectively, for the Twitter network~\cite{Myers2014} (the strongly connected component of  \textit{Swzone } and \textit{Psychlinks} is however, slighly smaller, around $30\%$ -- but see below). 
Furthermore, the diameters for the largest components are relatively small: the approximate diameter is $7$, and $90\%$ of all nodes are within $4$ hops of each other despite the graph's sparsity. 

Also, quotes are highly reciprocated: roughly $50\%$ of all node pairs connected by an arc sport an arc in the opposite direction, and $2-10\%$ of all triads are closed. The clustering coefficient, too, is remarkably high (above $0.3$); in particular, it remains high even for nodes of high degree, definitely more than in the Twitter or Facebook graphs~\cite{Myers2014} -- a possible explanation lying in the highly specialized nature of forums that tends to limit the variety of a user's circles. 

Finally, assortativity by node degree (informally, the propensity of nodes to link to nodes with roughly the same degree) is mildly positive for \textit{RPG} and \textit{Truemetal} (like in the Facebook or Twitter graphs \cite{Myers2014}), and mildly negative for \textit{Psychlinks} and \textit{Swzone} (as the Internet and WWW graphs \cite{Newman2003}). This finding is somewhat surprising, considering the lack of \textit{rich-get-richer} phenomena for users with high post count. An explanation might be that quoting follows social conventions different from simple posting and replying. The differences between \textit{RPG} and \textit{Truemetal}, and \textit{Swzone} and \textit{Psychlinks} match the intuition of the first pair of forums being driven by more "social", peer-to-peer conversations, and the second pair of forums being venues for obtaining information from experts.

\section{Friends and Fingerprints}\label{sec:applications}

In the previous section we saw how the forums' quote networks are structurally similar to modern, deliberate social networks. This section shows evidence that quote networks are indeed social in a more fundamental way. First of all, we demonstrate that quote usage is dependent on the identity of users, rather than just discourse. To do so, we use parameters of a user’s ego network (the induced subgraph including the node and all its neighbours) to identify him across distinct sets of discussions. Second, we show that quotes are good indicators of bonding between users. In fact, through the quote network we can recover most friendship relations established through the forums’ rarely used friendship mechanism.

\subsection{User fingerprinting}
Quotes are more than a tool for navigating group discussion - quoting patterns are characteristic of individual users, being in some sense weak digital fingerprints. This section shows that if we take a set of users, and partition their posts into two groups, it is possible to match the users in the two partitions comparing the quote networks built within each partition.

More precisely, we take a randomly chosen group of $n$ users, $n \in [2, 5, 10, 20, 50]$ (we only consider  users with at least $100$ posts, to remove noise). For each user in the group, we partition each of the threads he appears in, so that the total number of his posts in each partition is approximately balanced. Then, for each partition, we build the corresponding quote graph, using all quotes received from and made to the posts in the partition -- taking care to remove posts present in the other partition, if any (recall that quote graphs are directed graphs where users are nodes and edges are quote links between them, weighted by the actual number of quotes). For each of the $n$ users and for both graphs, we compute several network metrics characterizing the user’s ego network. The resulting feature vectors are then $L1$-normalized, after replacing missing values with the average value for the respective feature. We correctly identify a user if his feature vectors in the two partitions are the closest in terms of cosine similarity. We evaluate the identification algorithm using accuracy. We repeat the process $10$ times per forum, to stabilize results. The network metrics taken into consideration are reported in the list below.

\paragraph*{Author-quote network metrics for user fingerprinting}
\begin{itemize}\setlength\itemsep{0em}
\item[-]degree 
\item[-]in degree  
\item[-]out degree  
\item[-]self loops  
\item[-]number of triangles  
\item[-]clustering coefficient  
\item[-]square clustering coefficient  
\item[-]assortative mixing (all combinations of in and out degrees) 
\item[-]average neighbor degree  
\item[-]number of edges in ego network  
\item[-]number of nodes in ego network  
\item[-]ego network density  
\item[-]HITS: hubs,  authorities  
\item[-]pagerank  
\item[-]transitivity  
\item[-]eccentricity  
\item[-]vitality  
\item[-]closeness vitality 
\item[-]betweenness centrality  
\item[-]degree centrality  
\item[-]closeness centrality 
\item[-]katz centrality  
\item[-]communicability centrality  
\item[-]load centrality  
\item[-]eigenvector centrality  
\item[-]current flow betweenness centrality  
\item[-]current flow closeness centrality
\end{itemize}

Accuracy values exceed $80\%$ in all cases when attempting to  discriminate between two users, and decrease to around $30\%$ on average for $50$ users, considerably and consistently surpassing the random baseline in all forums. Results are also comparable to other approaches from the authorship attribution literature, where identification is performed analyzing the text of the users’ posts. The relatively lower accuracy for the swzone and psychlink forums may be due to their lesser adoption of quotes, which results in sparser, noisier networks. 

\begin{table}
\centering
\begin{tabular}[!htbp]{l c c c c}

 \textbf{\#users}   & \textbf{RPG}  & \textbf{SWZ}  & \textbf{TM}   & \textbf{PSY}  \\

 \textit{2}  & 1.00 & 0.85 & 0.80 & 0.85 \\

 \textit{5}  & 0.80 & 0.50 & 0.58 & 0.58 \\

 \textit{10} & 0.69 & 0.49 & 0.57 & 0.48 \\

 \textit{20} & 0.54 & 0.39 & 0.45 & 0.36 \\

 \textit{50} & 0.41 & 0.24 & 0.28 & 0.24 

\end{tabular}
\caption{Accuracy in user identification}
\label{tab:acc_ident}
\end{table}

\subsection{Friend prediction}
Many forums, while focusing mainly on discussion rather than networking, also provide simple affordances for letting users express their bonds within the forum’s community - akin to modern reciprocal social networks. A heuristic for evaluating a bond’s strength is observing the actual interaction that occurs between users. In this section, we build upon this idea, and deduce if two users are friends in the forum based on the author quote network.

\subsubsection{Friends in forums}
First, we present the friendship data for the four forums. A user can visit another user’s profile, and send a friendship request; if recipient accepts the request, the two users will be reciprocally shown in each other’s friends list. The friendship mechanism sees little use in all four forums (less than $10\%$ of all users), presumably because of its limited integration with the other services of the forums. If we analyze the friendship network, we see that most users who have at least one friend are connected through one or more degrees of separation (more than $90\%$ of nodes using friendship mechanisms are within a giant connected component, and more than $80\%$ of friendship edges are between nodes of that component). However, both the average degree of the friendship graph and the number of closed triangles in it are quite low, contrary to typical social data. The friendship networks are therefore rather sparse, and can be interpreted as a noisy subsample of the underlying community structure. Network statistics are reported in Table \ref{tab:quote_net}.

\begin{table}
\centering
\begin{tabular}[!htbp]{l c c c c}

                                 & \textbf{RPG}            & \textbf{SWZ} & \textbf{TM} & \textbf{PSY}     \\
\scriptsize
 Number of nodes                 & 3920           & 112    & 927       & 136            \\
\scriptsize
 Number of edges                 & 8040           & 232    & 2929      & 177            \\
\scriptsize
 Average degree                  & 4.1020         & 4.14 & 6.32    & 2.60         \\
\scriptsize
 Connected components            & 245            & 1      & 1         & 12             \\
\scriptsize
 Frac. nodes in largest cc & 0.85 & 1.0    & 1.0       & 0.84 \\
\scriptsize
 Frac. edges in largest cc & 0.95  & 1.0    & 1.0       & 0.94 \\
\scriptsize
 Diameter of largest cc          & 13             & 4      & 4         & 10             \\
\scriptsize
 Closed triangles                  & 4607           & 0      & 0         & 30             \\
\scriptsize
 Open triangles                     & 311335         & 3296   & 440410    & 1407           

\end{tabular}
\caption{Friend network statistics}
\label{tab:friend_net}
\end{table}

\subsubsection{Friendship and quotes}
We then try to gauge the relationship between the usage of quotes and the friendship system. We map users in the friendship network to users in the user quote network, and observe  how quote distribution changes in different types of relationships: when two users are friends, when they are not friends but both use the friendship system, when exactly one of the users uses the friendship system, and when neither of them does. Surprisingly, in the \emph{Swzone} dataset no quotes occur between friends, and in the \emph{Truemetal} dataset there are only two. However, the survival function of the multiplicities of quotes (the weight of the edges in the user quote graph, when made undirected) is more gradual and tends to reach higher multiplicities for friend users. This can be interpreted as more sociable users having relatively more prolonged quote exchanges than other users.

\subsubsection{Friend prediction}
Finally, we consider whether the user quote network can be leveraged to assess if two users are indeed friends in the forum. Note that this is a remarkably difficult task, considering 1) for both forums’ friendship mechanisms and social networks, it is questionable if their bonds are significant: for example, they are unable to differentiate mere acquaintances from best friends; 2) the ground truth for forum friends is scarce and noisy - it appears distant from a social phenomenon, as explained in the previous section.

We randomly sample $50$ pairs of friend users, and $50$ pairs where both users use the friendship system, but did not befriend one another. In particular, we obtain the latter via sampling $50$ users with a probability proportional to their degree in the friends graph, and sample a number of users that are not their friends, with probabilities again equal to their respective degree. We then build feature vectors using network metrics that are local to the nodes and their ego network, as well as metrics of co-occurrence in threads. The metrics are presented in the list below. Finally, we evaluate the accuracy of a LogisticRegression classifier in $10$ rounds of random partitioning into train and test set, maintaining an $80-20$ proportion and class balance. Accuracy appears around $70\%$ on average. This is very encouraging, considering that in two of the datasets friend users have no direct quotes between them. It is worth noting that if we frame the problem at a more local scale, and have to decide if two users posting \emph{in the same thread} are friends or not, the average accuracy rises above $80\%$. Accuracy results for all dataset are reported in Table \ref{tab:acc_friend}.

\begin{table}[!htbp]
\centering
\begin{tabular}[!htbp]{l c c c c}

 \textbf{sampling}   & \textbf{RPG}  & \textbf{SWZ}  & \textbf{TM}   & \textbf{PSY}  \\

 \textit{degree-based}  & 0.755 & 0.730 & 0.660 & 0.670 \\

 \textit{thread-based}  & 0.730 & 0.890 & 0.885 & 0.715 
\end{tabular}
\caption{Accuracy in friend prediction}
\label{tab:acc_friend}
\end{table}

\paragraph*{Author quote network metrics for friend prediction}
\begin{itemize}\setlength\itemsep{0em}
\item[-] number of directed edges in the pair
\item[-] number of common friends
\item[-] average clustering of common friends
\item[-] number of edges between common friends
\item[-] reciprocal of the fraction of edges that are not reciprocated
\item[-] reciprocity weighted by the out]degree of the nodes
\item[-] ratio of the minimum and the maximum of the edges in one direction among the pair
\item[-] fraction of the edges of the two nodes that are within the pair
\item[-] assortative mixing of the common friends
\item[-] minimum and maximum of the dispersion of the nodes in the pair
\item[-] minimum and maximum number of edges in one direction within the pair
\item[-] minimum and maximum of the average neighbor degrees for the nodes in the pair
\item[-] jaccard coefficient
\item[-] preferential attachment
\item[-] resource allocation index
\item[-] adamic adar index
\item[-] number of common threads
\item[-] jaccard index of the common threads
\item[-] delta measure on the number of authors in the common threads	
\item[-] adamic adar index on the number of authors in the common threads	
\item[-] sum of reciprocals of the number of authors in the common threads	
\item[-] product of the number of threads for both nodes in the pair
\end{itemize}

\section{Conclusions}\label{sec:conclusions}
Quotes in online forums are apparently simple tools, that nonetheless serve a variety of roles (from signals of common interest and acknowledgement, to aids for intra-thread navigation), and whose graph structure reveals a wealth of information both about forums and about individual posters. In particular, the quote graph provides each forum with a fingerprint that remains surprisingly invariant and accurate through many generations of users, and that can apparently distinguish between more ``social chat'' and more ``technical Q/A'' discussion venues. It can also identify individual users with fair accuracy, and uncover hidden social relationships even in the absence of mechanisms that we have come to think as fundamental to social networks, from followers/circles to likes and reputation mechanisms.

Quotes are ``higher resolution'' tools than likes, shares, and cites -- in this sense it would be interesting to see if the information they provide can still be recovered in networks that only offer tools of lower resolution. It would also be interesting to see if and how they can be used not only to identify users within a forum, but to identify users \emph{across} different forums, and to characterize their behaviours (distinguishing e.g. gurus from trolls). Finally, note that our work only looks at the graph structure of quotes and, somewhat tangentially, at their timing and length. This suggests on the one hand that a substantial portion of our analysis should be portable to contexts that do not involve text at all (e.g. image sharing venues), and at the same time that there is still much information to be uncovered by examining the actual quote text.

%
\bibliographystyle{abbrv}
\bibliography{sigproc}  
%
%
\balancecolumns 
\end{document}